\documentclass[preprint]{aastex}

\newcommand{\be}{\begin{equation}}
\newcommand{\ee}{\end{equation}}
\newcommand{\acon}{ {\cal A} } 
\newcommand{\tmod}{{ [t] }} 
\newcommand{\af}{{ \lambda }} 
\newcommand{\ybar}{{ \langle y \rangle }}  
\newcommand{\lamwig}{{ {\widetilde \lambda}_j }} 
\newcommand{\qwig}{{ {\widetilde q}_j }} 
\newcommand{\discrim}{{ \Delta }} 
\newcommand{\nstar}{{ N_\star }} 
\newcommand{\xihat}{{ {\hat \xi} }} 

\begin{document} 

\title{ORBITAL INSTABILITIES \\ 
IN A TRIAXIAL CUSP POTENTIAL} 

\author{Fred C. Adams,\altaffilmark{1,2} Anthony M. Bloch,\altaffilmark{1,3} \\
Suzanne C. Butler,\altaffilmark{1} Jeffrey M. Druce,\altaffilmark{1} and Jacob A. Ketchum\altaffilmark{1} } 

\altaffiltext{1}{Michigan Center for Theoretical Physics \\
Physics Department, University of Michigan, Ann Arbor, MI 48109} 

\altaffiltext{2}{Astronomy Department, University of Michigan, Ann Arbor, MI 48109} 

\altaffiltext{3}{Department of Mathematics, University of Michigan, Ann Arbor, MI 48109} 

\begin{abstract} 

This paper constructs an analytic form for a triaxial potential that
describes the dynamics of a wide variety of astrophysical systems,
including the inner portions of dark matter halos, the central regions
of galactic bulges, and young embedded star clusters.  Specifically,
this potential results from a density profile of the form $\rho (m)
\propto m^{-1}$, where the radial coordinate is generalized to
triaxial form so that $m^2 = x^2/a^2 + y^2/b^2 + z^2/c^2 $. Using the
resulting analytic form of the potential, and the corresponding force
laws, we construct orbit solutions and show that a robust orbit
instability exists in these systems. For orbits initially confined to
any of the three principal planes, the motion in the perpendicular
direction can be unstable.  We discuss the range of parameter space
for which these orbits are unstable, find the growth rates and
saturation levels of the instability, and develop a set of analytic
model equations that elucidate the essential physics of the
instability mechanism. This orbit instability has a large number of
astrophysical implications and applications, including understanding
the formation of dark matter halos, the structure of galactic bulges,
the survival of tidal streams, and the early evolution of embedded
star clusters.

\end{abstract} 


\section{INTRODUCTION} 

Many types of astrophysical objects are essentially collisionless
systems with extended mass distributions, including dark matter halos,
elliptical galaxies, galactic bulges, and star clusters.  Although
many of these systems display triaxial shapes, explicit analytic
models for the gravitational potentials, force laws, and orbits of the
triaxial incarnations of these systems are rare. This paper constructs
an analytic form for the triaxial potential of an important class of
systems, namely those in which the density profile approaches the form
$\rho \sim 1/r$ in the spherical limit. With the potential in hand, we
find the corresponding force terms (see also Poon \& Merritt 2001;
hereafter PM01) and study the orbits allowed by this triaxial system.
One contribution of this present work is the discovery of a robust
instability, wherein the orbital motion initially confined to any of
the three principal planes can be unstable to the growth of the
amplitude of the motion in the perpendicular direction.  In addition
to establishing the existence of the instability, this paper develops
a set of model equations that allow for a (mostly) analytic
description of the instability, the magnitude of the growth rate, and
the saturation mechanism.

This particular form for the density profile $\rho \sim 1/r$ arises in
many contexts. The Hernquist density profile, which has this form in
the inner limit (Hernquist 1990), was originally constructed to
describe the $R^{1/4}$ law in galactic bulges and elliptical galaxies.
Since the Hernquist profile arises in many other contexts, we
previously studied orbital solutions in the full (spherical) Hernquist
potential and obtained a number of analytic results that are
applicable to the present analysis (Adams \& Bloch 2005, hereafter
AB05). However, these bulge systems are known to be triaxial (Binney
\& Tremaine 1987, hereafter BT87; Binney \& Merrifield 1998), and
hence the triaxial generalization of the potential is important for
understanding these galactic-scale systems. For example, triaxial
potentials support box orbits, which allow stars to wander arbitrarily
close to the galactic center; this behavior affects the rate at which
stars interact with the central black hole.

On a larger scale, numerical simulations indicate that the density
profiles of dark matter halos display a nearly universal form in which
$\rho \sim 1/r$ in the inner regime (Navarro et al. 1997, hereafter
NFW). Recent numerical simulations that carry the calculations into
the future (Busha et al. 2005) show that the asymptotic form of the
density profile becomes steeper at large radii and is well-described
by a Hernquist profile (Hernquist 1990) but maintains the same $\rho
\sim 1/r$ form on the inside. Additional studies suggest that this
result for the inner form is robust; if the density profile is written
in the form $\rho \propto r^{-p}$, then $p \le 1.2$ (see Diemand et
al. 2004, Hayashi \& Navarro 2006, Huss et al. 1999, and references
therein).  Nonetheless, these systems are triaxial, rather than
perfectly spherically symmetric, with typical axis ratios of 5:4:3
(e.g., Kasun \& Evrard 2005, Jing \& Suto 2002). A host of other
authors have considered halo shapes, and always find roughly
ellipsoidal shapes, but considerable variation exists; in addition,
the axis ratios can vary with radius within the halo (see Allgood et
al. 2006 for further discussion).  Another result from numerical
simulations is that the inner portions of dark matter halos tend to
have (mostly) isotropic velocity distributions, while the outer
regions display more radial distributions; the orbit instability
considered here acts to isotropize the inner regions of these systems
and may help explain this finding.

On a smaller scale, the density profiles of cluster forming molecular
cloud cores are observed to have substantial non-thermal line-widths
(Larson 1985, Jijina et al. 1999); the size versus line-width relation
can be used to infer an effective equation of state, which, in turn,
implies a density profile of the form $\rho \sim 1/r$ in the spherical
limit (see Lizano \& Shu 1989; Jijina \& Adams 1996; McLaughlin \&
Pudritz 1996).  These cores often form small star clusters $(N = 30 -
3000)$ and the stellar members orbit within the corresponding
potential, which is dominated by the remaining gas (that has not
turned into stars).  These orbits determine the interaction rate
between young stellar objects and also determine their radiation
exposure (both processes generally have a destructive effect -- see,
e.g., Adams et al. 2006). These systems are also observed to be
triaxial, and roughly spheroidal, with typical axis ratios of order
2:1 (Myers et al. 1991, Ryden 1996). In addition, recent observations
suggest that some systems are highly flattened, with even more extreme
axis ratios (P. Myers, private communication).  The triaxial nature of
the potential allows star/disk systems to execute box orbits, which
brings them close to the cluster center where massive stars produce
large amounts of radiation (which can destroy the disks). On the other
hand, orbit instability acts to round out the clusters, even in the
absence of stellar scattering encounters. In any case, a complete
description of the dynamics of young embedded star clusters requires
an understanding of orbits in these triaxial systems.

Orbits and orbital instabilities in extended collisionless systems
have a long history. As is well known (BT87), triaxial systems display
a wide variety of orbits and a number of authors have studied orbits
in systems similar to those considered here (e.g., de Zeeuw \& Merritt
1983, Statler 1987; Holley-Bockelmann et al. 2001, 2002; PM01, Poon \&
Merritt 2002, Terzi{\'c} \& Sprague 2007, and many others).  The
radial orbit instability, where nearly radial orbits are unstable to
perpendicular motions, was first observed in numerical experiments by
Henon (1973).  This orbit instability was considered further by
several authors (including Barnes et al. 1986, Aguilar \& Merritt
1985, MacMillan et al. 2006).  The instability of motion in the
directions perpendicular to the principal planes was considered by
Binney (1981); subsequent work shows that this type of behavior arises
in triaxial potentials such as those considered herein (e.g., Merritt
\& Fridman 1996). However, a detailed (and analytic) description of
the instability mechanism has not been given and this issue represents
a main focus of this paper.

Study of instability in extended mass distributions can be carried out
in several ways: In one case, the global stability of the system is
considered and perturbation theory is used to study the evolution of
the distribution function (e.g., Aguilar \& Merritt 1985; Barnes et
al. 1986; Palmer \& Papaloizou 1987). Instead of considering the
distribution function, another approach is to consider the potential
as fixed and study the perturbations of the orbits themselves (e.g.,
Binney 1981, Scuflaire 1995, Papaphilippou \& Laskar 1998). Note that
these two types of instability are related, but are not equivalent.
If the distribution function is unstable, as in the former case, then
the orbits must change (and hence be unstable in a sense); it is
possible, however, for particular orbits to be unstable while the
distribution function remains stable.  In the context of individual
orbits, an important issue is to study the degree of chaos of
particular orbits, generally by computing the lyapunov exponents
(e.g., PM01, Valluri \& Merritt 1998). Although one expects more
chaotic orbits to lead to greater ``instability'', the two issues are
not equivalent. For example, orbits that are confined to a given plane
can be chaotic and even have large lyapunov exponents, but could
nonetheless remain stable to perturbations in the perpendicular
direction. The study of instability of individual orbits thus
represents a separate line of inquiry.  Here we adopt the latter
approach, follow the instability of individual orbits, and develop a
detailed description of the instability mechanism.

This paper builds on the aforementioned previous studies (see also de
Zeeuw \& Pfenniger 1988), and presents new results concerning the
instability of orbits in triaxial potentials.  We note that a great
deal of the previous work is either numerical or uses a simple
logarithmic model for the potential, where $\Phi \propto \ln [x^2/a^2
+ y^2 / b^2 + z^2 / c^2]$.  Although the numerical work cleanly
demonstrates the nature of the orbits and the existence of
instability, it does not elucidate the detailed mechanisms that lead
to the instability.  Previous analytic work using logarithmic
potential is limited to the study of axis ratios close to unity
because the corresponding density (e.g., obtained from $\nabla^2 \Phi
= 4 \pi G \rho$) becomes negative for more extreme axis ratios
(although more complicated generalizations allow for more extreme axis
ratios -- see Schwarzschild 1993).  In particular, the estimated
values of $a:b:c$ = 5:4:3 for dark matter halos (Kasun \& Evrard 2005)
are over the limit for the applicability of the usual logarithmic
model potential. For example, if we scale the axes so that $a=1$, the
requirement for the logarithmic potential to have nonnegative density
everywhere can be written in the form $c > b/\sqrt{1+b^2}$. For $b =
4/5$, typical of dark matter halos, this constraint becomes $c >
4/\sqrt{41} \approx 0.62$ (i.e., the typical the axis ratios found in
numerical simulations just fail to satisfy this constraint).  For
$b=1/2$, typical of cluster forming molecular cloud cores, the
constraint is more restrictive, $c \ge 1/\sqrt{5} \approx 0.45$ (just
below $b$ = 1/2), whereas we expect even more extreme axis ratios
(perhaps as small as $c \sim 0.10$).

Although the results of this work are applicable to a wide range of
astronomical problems (outlined above), this paper focuses on the
construction of the triaxial potential, a brief description of its
orbits, and a detailed analysis of the orbit instability. The paper is
organized as follows. In \S 2 we construct the analytic form for the
potential resulting from the triaxial generalization of a density
profile $\rho \sim 1/r$, and present the corresponding analytic
expressions for the forces and axis ratios.  In \S 3 we briefly survey
the possible orbit solutions in the triaxial potential.  Next we study
the stability of orbits, and show that orbits in all three principal
planes can be unstable to motion in the perpendicular direction (\S
4); we also develop analytic model equations to describe the
instability mechanism (\S 5). The paper concludes, in \S 6, with a
summary of results and a discussion of their astrophysical
implications. Although this paper focuses on the inner limit (where
$\rho \sim 1/r$), Appendix A presents analytic expressions for the
potential and force terms valid over the full radial extent of the
axisymmetric version of the NFW profile.

\section{ANALYTIC POTENTIAL AND FORCES} 

The overarching goal of this study is to understand orbits in the
potentials resulting from a generalized density profile of the form
\be 
\rho_{\rm gen} = \, \rho_0 \, {f(m) \over m} \, , 
\label{eq:rhogen} 
\ee 
where $\rho_0$ is a density scale and 
where the variable $m$ has a triaxial form 
\be
m^2 = {x^2 \over a^2} + {y^2 \over b^2} + {z^2 \over c^2} \, . 
\label{eq:mdef} 
\ee 
Keep in mind that the variable $m$ defines the surfaces of constant density, 
whereas the radial coordinate $\xi$ is given by 
\be
\xi^2 = x^2 + y^2 + z^2 \, .
\label{eq:xidef} 
\ee
The function $f(m)$ is assumed to approach unity as $m \to 0$ so that
the density profile has the form $\rho \sim 1/m$ in this inner limit.
Two standard forms for extended mass distributions (with many
astrophysical applications) are the NFW profile (Navarro et al. 1997)
and the Hernquist profile (Hernquist 1990), which have density
profiles of the forms
\be 
\rho_{\rm nfw} = {\rho_0 \over m (1 + m)^2} \qquad {\rm and} \qquad 
\rho_{\rm hq} = {\rho_0 \over m (1 + m)^3} \, . 
\ee 
For the rest of this analysis, we use a dimensionless formulation so
that $\rho_0$ = 1. In the spherical limit, $m = r/r_s$, where $r_s$ is
a scale radius (NFW).  In this treatment, we normalize all length
scales by $r_s$ to make the variables $(x,y,z)$ and the geometric
constants $(a,b,c)$ dimensionless.

For any density profile of this general (triaxial) form, the 
potential can be found by evaluating the integral 
\be 
\Phi (x,y,z) = 2 \int_0^\infty {\psi(m) du \over (a^2 + u)^{1/2} 
(c^2 + u)^{1/2} (b^2 + u)^{1/2} } \, , 
\label{eq:phibasic} 
\ee 
where $m$ is given by equation (\ref{eq:mdef}) 
where the function $\psi(m)$ is defined through the relation (BT87, 
Chandrasekhar 1969) 
\be 
\psi(m) = \int_m^\infty 2 m dm \rho (m) \, . 
\ee
We then define a constant $\acon$ by 
\be 
\acon \equiv 2 \int_0^\infty f(m) dm \, , 
\ee 
where we assume that the integral converges, so that the function
$\psi$ takes the form $\psi = \acon - 2 m$ in the inner limit ($m \to
0$).  For the NFW (Hernquist) profile, the integral is easily evaluated
and the constant $\acon$ = 2 (1). Keep in mind that in this
dimensionless formulation, we have set a number of constants to unity
and let the potential have a positive value.

\subsection{Axisymmetric Inner Limiting Form} 

For the axisymmetric problem, we can set $a$ = 1 = $b$ and define $R^2
\equiv x^2 + y^2$.  After defining two new ``constants'' according to
\be 
\gamma \equiv 1 -  c^2 \qquad {\rm and} \qquad q \equiv \gamma R^2/\xi^2 \, , 
\ee
the potential in the inner limit can be written in the form 
\be 
\Phi \approx 2 \acon \gamma^{-1/2} \cos^{-1} (c) + 
{2 z \over \gamma} \ln \Big| { (\xi + z) (\xi \sqrt{1-q} - z) \over 
(\xi - z) (\xi \sqrt{1-q} + z) } \Big| + {4 \xi \over \gamma} 
(\sqrt{1 - q} - 1) \, ,  
\label{eq:axial} 
\ee
where the constant $\acon$ depends on the form of the density profile
over the rest of its range (as defined above). We note that for the
axisymmetric version of the NFW profile, the full potential (valid for
all $m$, not just in the inner limit) can be found analytically. This
result is presented in Appendix A, and equation (\ref{eq:axial})
agrees with the inner limit of the general NFW potential.

\subsection{Triaxial Inner Limiting Form} 

Next we generalize to the triaxial case, where $a > b > c > 0$. 
If we consider the inner limit of a density profile of the form of 
equation (\ref{eq:rhogen}), then the potential can be written as 
two terms
\be
\Phi = \acon I_1 - 2 I_2 \, , 
\label{eq:phistart} 
\ee 
where $\acon$ has the same meaning as before and where 
\be
I_1 \equiv \int_0^\infty {du \over \sqrt{a^2 + u} 
\sqrt{b^2 + u} \sqrt{c^2 + u} } \, , 
\ee
and 
\be
I_2 \equiv \int_0^\infty \, du \, 
{ \bigl[ \xi^2 u^2 + \Lambda u + \Gamma \bigr]^{1/2} \over 
(a^2 + u) (b^2 + u) (c^2 + u) } \, . 
\ee 
The radial coordinate $\xi$ is defined through equation
(\ref{eq:xidef}) and the remaining coefficients in the numerator 
are given by the following functions of the coordinates,
\be 
\Lambda \equiv (b^2+c^2) x^2 + (a^2+c^2) y^2 + (a^2+b^2) z^2 
\qquad {\rm and} \qquad 
\Gamma \equiv b^2 c^2 x^2 + a^2 c^2 y^2 + a^2 b^2 z^2 \, . 
\label{eq:gamdef} 
\ee 

Note that the first integral $I_1$ defines the depth of the potential
well.  Further, $I_1$ does not depend on the spatial coordinates ---
it is a constant for a given set of axis ratios $(a, b, c)$ --- and is
determined by the obliquity parameter $\eta_{ob} \equiv (a^2 - b^2) / 
(a^2 - c^2)$.

The second integral $I_2$ can be expanded into three terms according to
$$
I_2 =   {1 \over (a^2 - b^2) (a^2 - c^2)} \int_0^\infty du 
{ \bigl[ \xi^2 u^2 + \Lambda u + \Gamma \bigr]^{1/2} \over 
a^2 + u} \, \qquad \qquad \, 
$$
$$
\qquad \qquad  -  
{1 \over (a^2 - b^2) (b^2 - c^2)} \int_0^\infty du 
{ \bigl[ \xi^2 u^2 + \Lambda u + \Gamma \bigr]^{1/2} \over 
b^2 + u} \, 
$$
\be 
\qquad \qquad \qquad \qquad \qquad \qquad \qquad \qquad 
+  {1 \over (a^2 - c^2) (b^2 - c^2)} \int_0^\infty du 
{ \bigl[ \xi^2 u^2 + \Lambda u + \Gamma \bigr]^{1/2} \over 
c^2 + u} \, .  
\ee
The integrals can be evaluated individually and recombined 
to take the form 
$$
I_2 = {1 \over (a^2 - b^2) (a^2 - c^2) (b^2 - c^2)} \Biggl\{ 
F(a) (c^2 - b^2) \ln \Bigl[ {\Lambda a^2 - 2 \xi^2 a^4 + 2 F(a) \xi a^2 
\over 2 \Gamma - \Lambda a^2 + 2 F(a) \sqrt{\Gamma} } \Bigr] + 
$$
\be
F(b) (a^2 - c^2) \ln 
\Bigl[ {\Lambda b^2 - 2 \xi^2 b^4 + 2 F(b) \xi b^2 
\over 2 \Gamma - \Lambda b^2 + 2 F(b) \sqrt{\Gamma} } \Bigr] + 
F(c) (b^2 - a^2) \ln 
\Bigl[ {\Lambda c^2 - 2 \xi^2 c^4 + 2 F(c) \xi c^2 
\over 2 \Gamma - \Lambda c^2 + 2 F(c) \sqrt{\Gamma} } 
\Bigr]  \Biggr\} \, , 
\label{eq:final} 
\ee
where the function $F$ is defined by 
\be 
F(\alpha) \equiv \big[ \xi^2 \alpha^4 - \Lambda \alpha^2  
+ \Gamma \bigr]^{1/2} \, . 
\ee 
The spatial dependence of the potential is contained in the functions
$\xi$, $\Lambda$, and $\Gamma$, which depend on the usual coordinates
$(x,y,z)$ through the relations of equation (\ref{eq:gamdef}). Notice
that the second term, as written, has both real and imaginary parts,
since $F^2(b)$ is negative (under the usual ordering $a > b > c$). We 
get agreement between the analytic form and the numerically evaluated
form when we take the real part of equation (\ref{eq:final}).
Alternatively, we can rewrite the second term so that the potential
integral takes the form
$$
I_2 = {1 \over (a^2 - b^2) (a^2 - c^2) (b^2 - c^2)} \Biggl\{ 
F(a) (c^2 - b^2) \ln \Bigl[ {\Lambda a^2 - 2 \xi^2 a^4 + 2 F(a) \xi a^2 
\over 2 \Gamma - \Lambda a^2 + 2 F(a) \sqrt{\Gamma} } \Bigr]  
\, \qquad \qquad \, \qquad \, 
$$
$$
+ |F(b)| (a^2 - c^2) \Bigl[ \sin^{-1} \Bigl( 
{\Lambda - 2 b^2 \xi^2 \over \sqrt{\Lambda^2 - 4 \xi^2 \Gamma} } 
\Bigr) - \sin^{-1} \Bigl( 
{2 \Gamma/b^2 - \Lambda \over \sqrt{\Lambda^2 - 4 \xi^2 \Gamma} } 
\Bigr) \Bigr] 
$$
\be
\, \qquad \qquad \qquad \, \qquad \qquad + F(c) (b^2 - a^2) \ln  
\Bigl[ {\Lambda c^2 - 2 \xi^2 c^4 + 2 F(c) \xi c^2 
\over 2 \Gamma - \Lambda c^2 + 2 F(c) \sqrt{\Gamma} } 
\Bigr]  \Biggr\} \, ,   
\label{eq:finaltwo} 
\ee
where the second term is now manifestly real.  We also note that in
the evaluation of the second term, it is sometimes advantageous to use
a trigonometric identity to write the $\sin^{-1}$ terms as $\tan^{-1}$
expressions.

\subsection{Force Terms} 

The components of the force can be obtained by direct differentiation
of the potential (eq. [\ref{eq:finaltwo}]). Alternatively, one can
start with the original integral expression for the potential,
differentiate first, and then perform the integration. The second
procedure is simpler, but both result in force components of the
following forms (where we have re-introduced the minus sign): 
\be
{\cal F}_x = - 2 {d I_2 \over dx} = - {2 x \over F(a)} 
\ln \Bigg| {2 F(a) \sqrt{\Gamma} + 2 \Gamma - \Lambda a^2 \over 
a^2 \bigl[ 2 F(a) \xi + \Lambda  - 2 a^2 \xi^2 \bigr] }
\Bigg| \, , 
\label{eq:xforce} 
\ee
\be
{\cal F}_y = - 2 {d I_2 \over dy} = - {2 y \over |F(b)|} 
\Bigl[ \sin^{-1} \Bigl( {\Lambda - 2 b^2 \xi^2 \over 
\sqrt{\Lambda^2 - 4 \xi^2 \Gamma} } \Bigr) - \sin^{-1} \Bigl( 
{2 \Gamma/b^2 - \Lambda \over \sqrt{\Lambda^2 - 4 \xi^2 \Gamma} } 
\Bigr) \Bigr] \, , 
\label{eq:yforce} 
\ee
\be 
{\cal F}_z = - 2 {d I_2 \over dz} = - {2 z \over F(c)} 
\ln \Bigg| {2 F(c) \sqrt{\Gamma} + 2 \Gamma - \Lambda c^2 \over 
c^2 \bigl[ 2 F(c) \xi + \Lambda  - 2 c^2 \xi^2 \bigr] }
\Bigg| \, . 
\label{eq:zforce} 
\ee
We note that equivalent expressions for the force terms have been
derived previously (PM01), and that the results can be expressed in a
variety of forms. In particular, the $\sin^{-1}$ functions can be
written as $\tan^{-1}$ functions; for example, in ther first term of
equation (\ref{eq:yforce}), $\Theta \equiv \sin^{-1} \bigl[ (\Lambda -
2 b^2 \xi^2) / \sqrt{\Lambda^2 - 4 \xi^2 \Gamma} \bigr] = \tan^{-1}
\bigl[ (\Lambda - 2 b^2 \xi^2) / 2 \xi \sqrt{\Lambda b^2 - \Gamma -
b^2} \bigr]$.  Both forms are useful for numerical evaluation of the
forces, depending on the context. In addition, the leading coefficients 
of the force terms obey identities that allow for different expressions,
e.g., $2x/F(a) = 2 (a^2 - b^2)^{-1/2} (a^2 - c^2)^{-1/2}$, with similar 
forms for the other forces. 

\subsection{Shape of Equipotential Contours} 

One common feature of triaxial systems is that the surfaces of
constant potential are generally rounder (closer to spherical
symmetry) than the surfaces of constant density, which are ellipsoidal
(by construction) for the class of systems considered here. In these
systems, the surfaces of constant potential are given by setting
equation (\ref{eq:finaltwo}) equal to a constant; although this form
is analytic, it remains both implicit and algebraically complicated.
However, we can find the ``axis ratios'' for the potential surfaces by
evaluating the potential along each of the principal axes. For
example, along the $x$-axis, the potential reduces to the form
\be
\Phi = \Phi_0 - x J_x(a,b,c) \, , 
\label{eq:phiaxis} 
\ee
where all of the geometry is encapsulated in the function 
\be 
J_x = {2 \over (a^2 - c^2)^{1/2}  (a^2 - b^2)^{1/2} } \,  
\ln \Bigg| {b (a^2 - c^2)^{1/2} - c (a^2 - b^2)^{1/2} \over 
a \bigl[ (a^2 - c^2)^{1/2} - (a^2 - b^2)^{1/2} \bigr] } \Bigg| \, . 
\label{eq:jxint} 
\ee 
Along the other two axes, the potential takes a similar form with 
\be
J_y = {1 \over (a^2 - b^2)^{1/2}  (b^2 - c^2)^{1/2} } \,  \Bigl\{ 
\sin^{-1} \Bigl[ {a^2 + c^2 - 2 b^2 \over a^2 - c^2} \Bigr] - 
\sin^{-1} \Bigl[ {a^2 (c^2 - b^2) + c^2 (a^2 - b^2 ) \over a^2 - c^2 } 
\Bigr] \Bigr\} \, , 
\label{eq:jyint} 
\ee
and 
\be 
J_z = {- 2 \over (a^2 - c^2)^{1/2}  (b^2 - c^2)^{1/2} } \,  
\ln \Bigg| {b (a^2 - c^2)^{1/2} - a (b^2 - c^2)^{1/2} \over 
c \bigl[ (a^2 - c^2)^{1/2} - (b^2 - c^2)^{1/2} \bigr] } \Bigg| \, . 
\label{eq:jzint} 
\ee 
The surfaces of constant density are ellipsoids. For example, if we
want to find the locations on the $x$-axis and the $z$-axis with the
same density, we require $x/a = z/c$ or $z/x = c/a$ (where we take
positive values).  Proceeding in analogous fashion, if we want to find
the locations on the $x$-axis and the $z$-axis with the same value of
the potential, we require $x J_x = z J_z$ so that $z/x = J_x/J_z$. The
other axis ratios can be found similarly. Further, the above equations
imply that $c/a < J_x(a,b,c)/J_z(a,b,c)$, and similarly for the other
pairs of axes, so that the surfaces of constant potential are indeed
rounder (closer to spherical) than the surfaces of constant density. 
Moreover, we have obtained analytic expressions for the axis ratios. 

In the limit of extreme axis ratios, the above expressions simplify
considerably. In the limit $a \gg 1$ and $c \ll 1$, the integrals that
define the shape of the potential approach the asymptotic forms
\be
J_x \to {2 \ln (2a) \over a^2} \, , \qquad 
J_y \to {\pi \over a} \, , \qquad {\rm and} \qquad 
J_z \to {2 \ln (2/c) \over a} \, . 
\ee 
In the opposite limit, where $a \to 1$ and $c \to 1$, the potential
becomes spherical. 

\subsection{Nearly Spherical Density Profiles} 

In some applications, it is useful to have the form of the potential
for a density distribution that is nearly spherical, but nonetheless
displays triaxial departures from perfect symmetry. For the sake of
definiteness, this subsection considers the triaxial density profile
with axis ratios
\be 
a^2 = 1 + \delta_a \qquad b=1, \qquad 
{\rm and} \qquad c^2 = 1 - \delta_c \, , 
\ee
where physically meaningful solutions require $0 < \delta_a, \delta_c
< 1$, but this form is most useful in the limit $\delta_a, \delta_c
\ll 1$. After propagating this ansatz through the equations developed
above, the potential can be written in the relatively simple form
\be 
\Phi = \Phi_0 - {\xi \over 2} - 
{ ( \delta_c z^2 - \delta_a x^2) \over 8 \xi } \, , 
\label{eq:phidelt} 
\ee 
where $\xi$ is the (usual) radial coordinate given by equation
(\ref{eq:xidef}).  The density corresponding to this potential 
takes the form
\be 
\rho = {1 + 2 (\delta_c - \delta_a) \over \xi} + 
{ (\delta_a x^2 - \delta_c z^2) \over 2 \xi^3 } \, . 
\label{eq:rhodelt} 
\ee 
It is straightforward to verify that this form is the leading order
correction to the general density profile of equation 
(\ref{eq:rhogen}) in the limit of small departures from
sphericity. Further, this density field will be positive as long as
$\delta_a < 1/2 + 3 \delta_c/4$, i.e., for sufficiently spherical
systems. For more flattened profiles, the full form of the equations
presented in the previous sections must be used. Note that this
potential-density pair has a density profile of the form $\rho \sim
1/\xi$ in the spherical limit, rather than the more commonly used
logarithmic potential, which corresponds to a density profile of the
form $\rho \sim 1/\xi^2$. As a result, this form provides a useful
alternative to the logarithmic potential and can be applied to many
astrophysical systems of interest.  Notice also that the
density/potential pair found above simplifies even further for the
symmetric case where $\delta_c$ = $\delta_a$.

\section{ORBITS} 

A host of previous authors have studied the various orbits that are
supported by triaxial potentials such as those considered herein
(e.g., see BT87, Richstone 1982, Hunter \& de Zeeuw 1992;
Schwarzschild 1993, PM01, Holley-Bockelmann et al. 2002). A wide
variety of such orbits exist, including radial orbits, tube orbits,
box orbits, and resonant orbits. A full accounting and presentation of
all of the possible orbits has (mostly) been covered by the
aforementioned previous work and is not the goal of this
paper. Instead, we are primarily concerned with studying the orbit
instability that arises in these systems, as well as obtaining
analytic results whenever possible.  This section develops analytic
solutions for principal axis orbits (\S 3.1) and for radial orbits in
the spherical limit (\S 3.2). A brief discussion of the surfaces of
section, the fraction of box orbits versus loop orbits, and the
dependence of these results on the axis ratios is presented in \S 3.3.

Throughout this paper, when numerical integration of the orbits is
required, we use a Bulirsch-Stoer (B-S) integration scheme (Press et
al. 1992). This method is both accurate and explicit. For the systems
at hand, our B-S scheme incurs errors in relative accuracy of order 1
part in $10^{11}$ per total time step, with typical accumulated errors
of 1 part in $10^9$, small enough that numerical error is not an
issue. For completeness, we note that some longer runs accumulate a
total error of 1 part in $10^7$.

\subsection{Principal Axis Orbits}

For orbits that are constrained to one of the principal axes, the
equation of motion reduces to the simple form
\be 
{d^2 w_j \over dt^2 } = - g_j = {constant} \, , 
\ee
where $w_j$ represents the $x$, $y$, or $z$ coordinate. The equation
is thus the same as that of a baseball thrown vertically in the
Earth's gravitational field (with no air resistance), although the
trajectories repeat (oscillate) in the present application. The
constant force strengths follow directly from equations
(\ref{eq:xforce} -- \ref{eq:zforce}).

For orbits along the principal axes, the force terms are constant 
and can most easily be written in integral form
\be
g_x = \int_0^\infty {du \over (u + a^2) 
\bigl[ (u + b^2) (u + c^2) \bigr]^{1/2} } \, , 
\ee 
\be
g_y = \int_0^\infty {du \over (u + b^2) 
\bigl[ (u + a^2) (u + c^2) \bigr]^{1/2} } \, , 
\ee 
\be
g_z = \int_0^\infty {du \over (u + c^2) 
\bigl[ (u + a^2) (u + b^2) \bigr]^{1/2} } \, .  
\ee

These integrals can be evaluated. For example, the force along the 
x-axis can be written 
\be 
g_x = {1 \over \sqrt{\gamma_x} }  \ln \Biggl| 
{2 [  \gamma_x + bc \sqrt{\gamma_x} ]/a^2 + b^2 + c^2 - 2 a^2 
\over 2 \sqrt{\gamma_x} + b^2 + c^2 - 2 a^2 } \Biggr| \, , 
\ee 
where we have defined $\gamma_x \equiv (a^2-b^2)(a^2 -c^2)$. 
As before, the integral along the intermediate axis takes a 
different form, 
\be
g_y = {1 \over \sqrt{\gamma_y} }  \Biggl\{ \sin^{-1} 
\Bigl[ {a^2 + c^2 - 2 b^2 \over a^2 - c^2 } \Bigr] - 
\sin^{-1} \Bigl[ {a^2 + c^2 - 2 b^2 + 2 \sqrt{\gamma_y} / 
c^2 \over a^2 - c^2 } \Bigr] \Biggl\} \, , 
\ee
where we define $\gamma_y \equiv (a^2-b^2)(b^2 -c^2)$, 
whereas the integral along the shortest axis becomes 
\be 
g_z = {1 \over \sqrt{\gamma_z} }  \ln \Biggl| 
{2 [  \gamma_z + ab \sqrt{\gamma_z} ]/c^2 + a^2 + b^2 - 2 c^2 
\over 2 \sqrt{\gamma_z} + a^2 + b^2 - 2 c^2 } \Biggr| \, , 
\ee 
where $\gamma_z \equiv (a^2-c^2)(b^2 -c^2)$. 

\subsection{Radial Orbits in the Spherical Limit} 

In the limit of a spherical Hernquist potential, 
the radial equation of motion takes the simple form 
\be 
{d^2 \xi \over dt^2} + {\omega_0^2 \over (1 + \xi)^2} 
= 0 \, ,
\ee
where $\xi=r/r_s$ is the dimensionless radial coordinate 
and where 
\be
\omega_0^2 \equiv 2 \pi G \rho_0 \, . 
\ee 
The parameter $\omega_0$ thus sets the fundamental time 
unit for this potential. This equation of motion can 
be directly integrated to find the solution. For the 
case of an orbit starting with initial radial coordinate 
$\xi_0$ and zero (radial) velocity, the solution takes 
the form 
\be 
{\sqrt{2} \omega_0 t \over (1 + \xi_0)^{3/2} } = 
\cos^{-1} z + z \sqrt{1 - z^2} \, , 
\label{eq:radsphere} 
\ee 
where the radial coordinate is written in terms of 
the variable $z$, which is defined by 
\be
z^2 \equiv {1 + \xi \over 1 + \xi_0} \, . 
\ee 
Note that equation (\ref{eq:radsphere}) determines the period of a
radial orbit for a given starting point $\xi_0$; specifically, the
half-period is given by
\be 
\tau_{1/2} = {1 \over \sqrt{2} \omega_0} \Bigl[ 
(1 + \xi_0)^{3/2} \cos^{-1} \sqrt{1 + \xi_0} + 
\sqrt{\xi_0 + \xi_0^2} \Bigr] \, . 
\label{eq:periodfull} 
\ee
In the inner limit where $\xi \ll 1$, the solution of equation
(\ref{eq:radsphere}) reduces to the solution of a constant force
equation and the time scale of a half period (the time required for on
inward or outward crossing) simplifies to the form 
\be
\tau_{1/2} = {\sqrt{2 \xi_0} \over \omega_0} \, . 
\label{eq:periodin} 
\ee 
For general (non-radial) orbits, the half-periods and turning angles
have been studied previously for the case of a spherical Hernquist
potential (AB05).

\subsection{Brief Survey of Orbits} 

\begin{figure}
\figurenum{1a}
{\centerline{\epsscale{0.90} \plotone{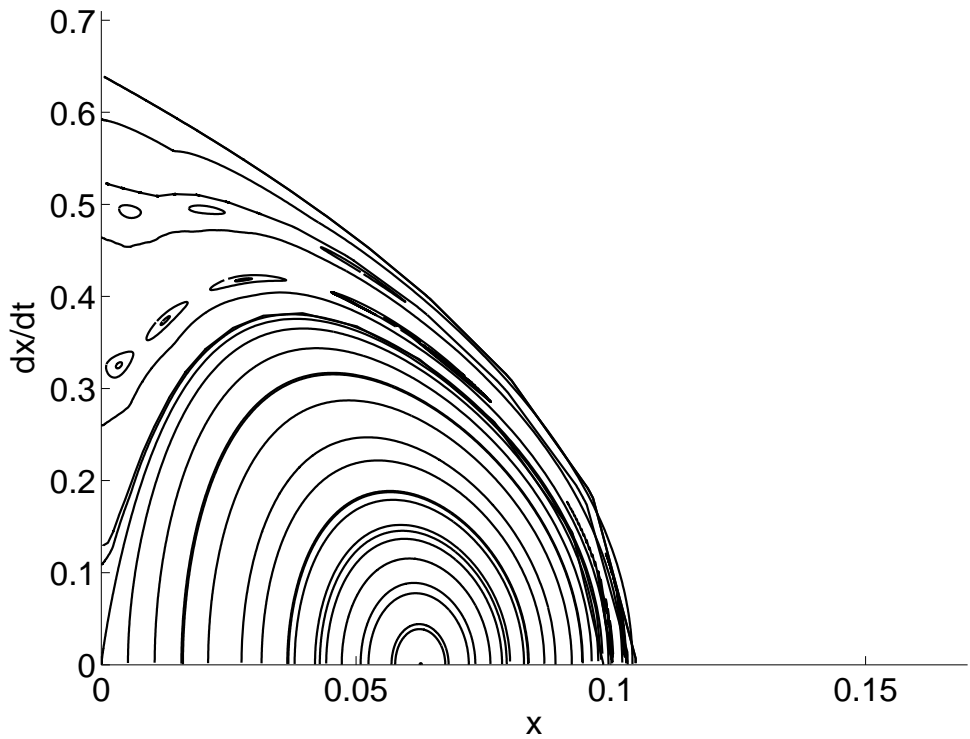} }}
\figcaption{Surfaces of section for triaxial potential with varying
axis weights.  (a) Surface of section for nearly spherical version of
the potential with $(a,c)$ = $(\sqrt{1.1},\sqrt{0.9})$.  (b) Surface
of section for moderate axis weights $(a,c)$ = $(1.25,0.75)$, so that
the axis ratios are 3:4:5. (c) Surface of section for elongated
potential with $(a,c) = (\sqrt{3},\sqrt{2}/2)$. The letters refer to
the orbits shown in Figure \ref{fig:gallery}. }  
\label{fig:section1} 
\end{figure}

The triaxial system considered herein supports a wide variety of
orbits, and displays typical dynamical behavior for this class of
astrophysical system (BT87).  Previous work shows that as the axis
ratios increase (i.e., as the ellipsoidal density contours become more
elongated), the fraction of orbits that are chaotic increases (PM01).
Similarly, as the axis ratios increase, the fraction of box orbits
increases and the fraction of loop orbits decreases (BT87). Both of
these trends are illustrated in Figure 1, which shows a series of
surfaces of section with increasing axis ratios. For all three
systems, the maximum value of the triaxial coordinate $m$ is taken to
be $m_{\rm max} = 0.1$; the maximum value of the $x$-coordinate is
then given by $x_{\rm max} = a m_{\rm max}$ (where $a$ is the value of
the axis weight).  The energy measured relative to the minimum value
at the center of the potential well is nearly constant (with values
$\Delta E$ = 0.205, 0.224, and 0.216 for the cases shown). Since we
are primarily interested in the instability of orbits that are
initially confined to one of the principal planes, these orbits are
all fixed in the $x-z$ plane; similar results hold for orbits started
in the other two planes.  The first plot in Figure \ref{fig:section1}
uses axis weights near unity so that the mass distribution is nearly
spherical. As expected (BT87), the surface of section for this system
is well ordered and most of the phase plane corresponds to loop
orbits. Note that loop orbits correspond to curves in the surface of
section where the end points terminate along the positional ($x$)
axis, whereas the box orbits have patterns in the plane that continue
into the velocity (vertical) axis.  As the axis weights depart farther
from unity (see Fig. \ref{fig:section1}), the surface of section plot
breaks up into an increasing number of islands and contains a lower
fraction of loop orbits.

{\centerline{\epsscale{0.90} \plotone{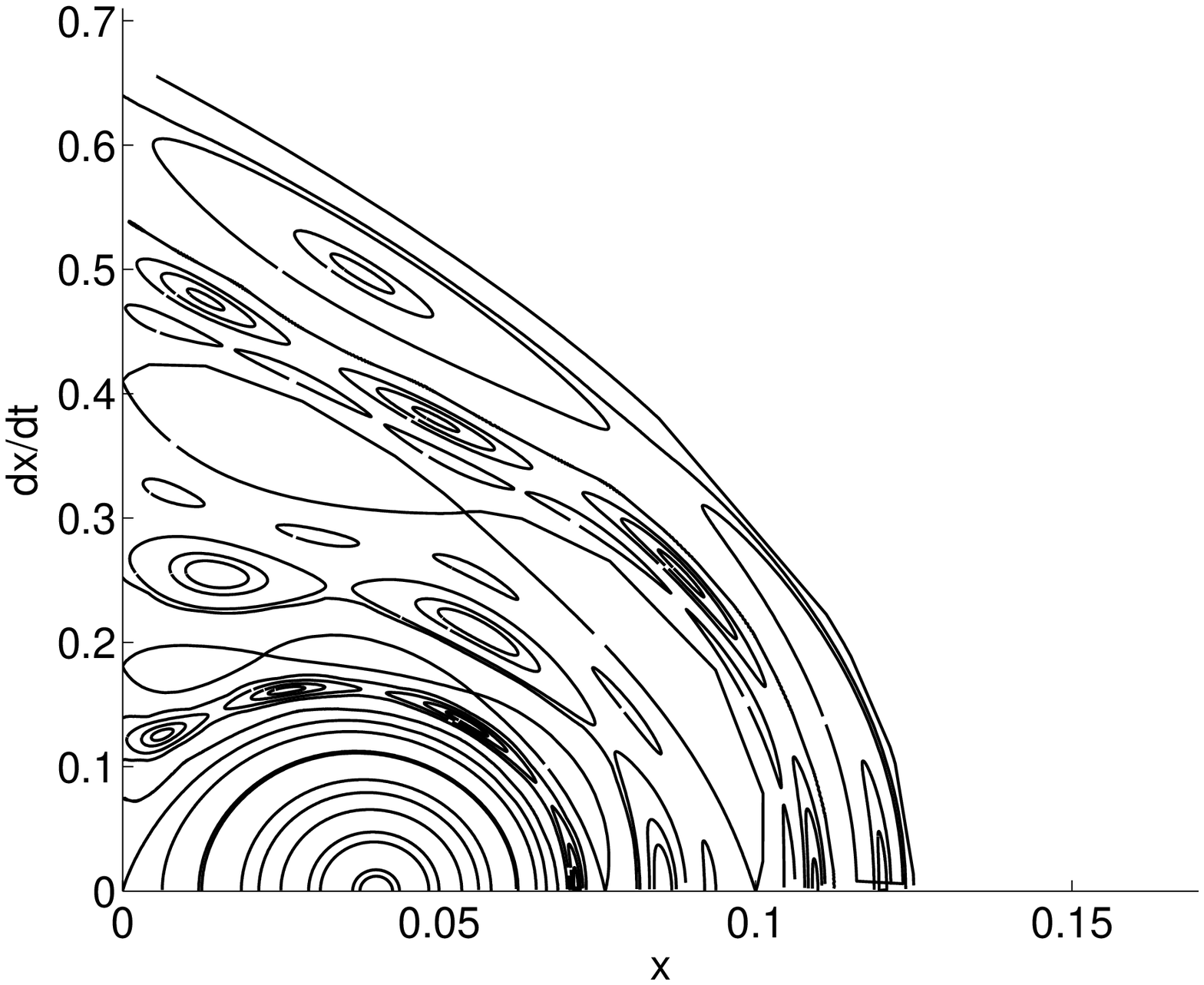} }}
\centerline{Fig. 1 --- Continued.} 

{\centerline{\epsscale{0.90} \plotone{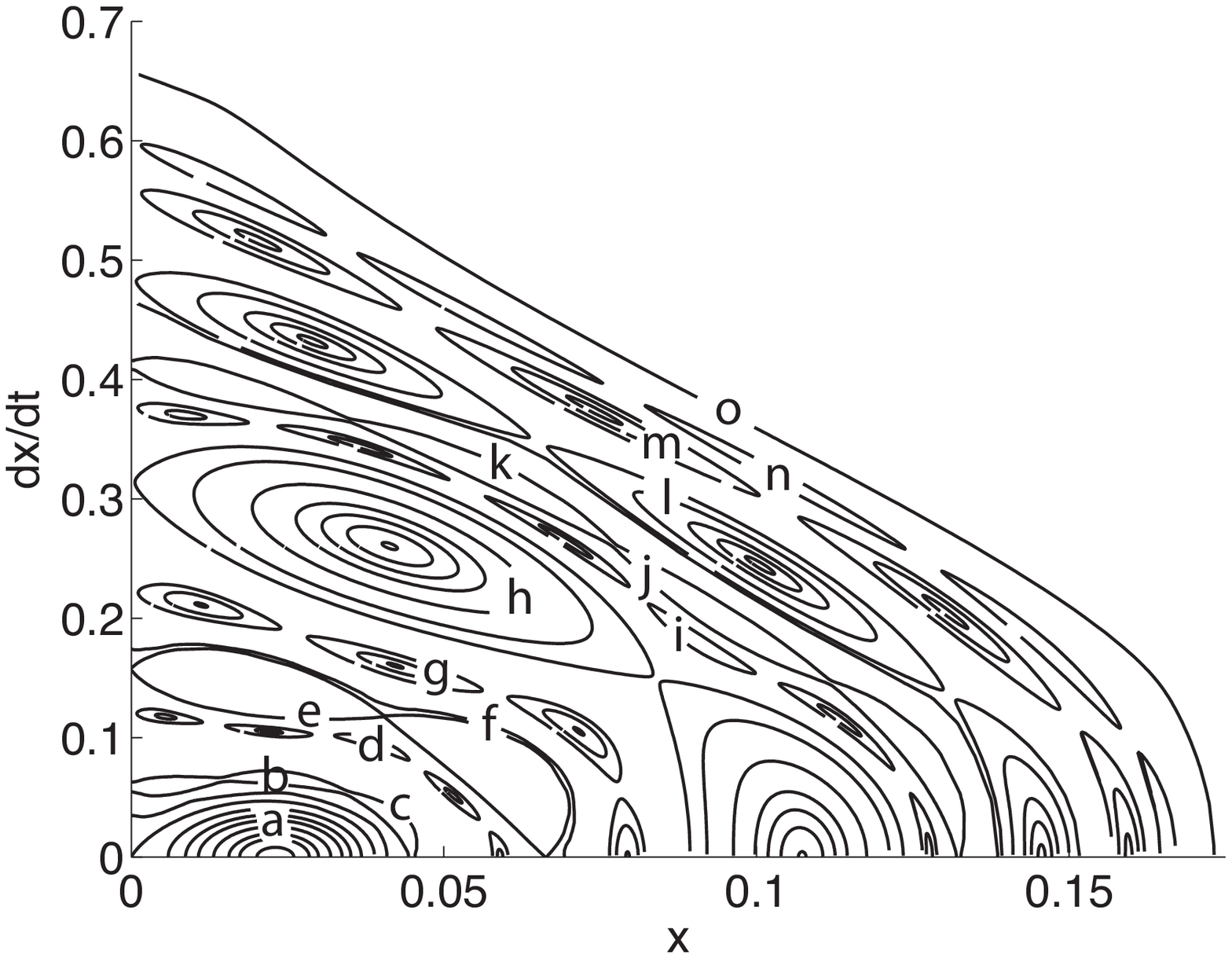} }} 
\centerline{Fig. 1 --- Continued.} 


\begin{figure}
\figurenum{2}
{\centerline{\epsscale{0.80} \plotone{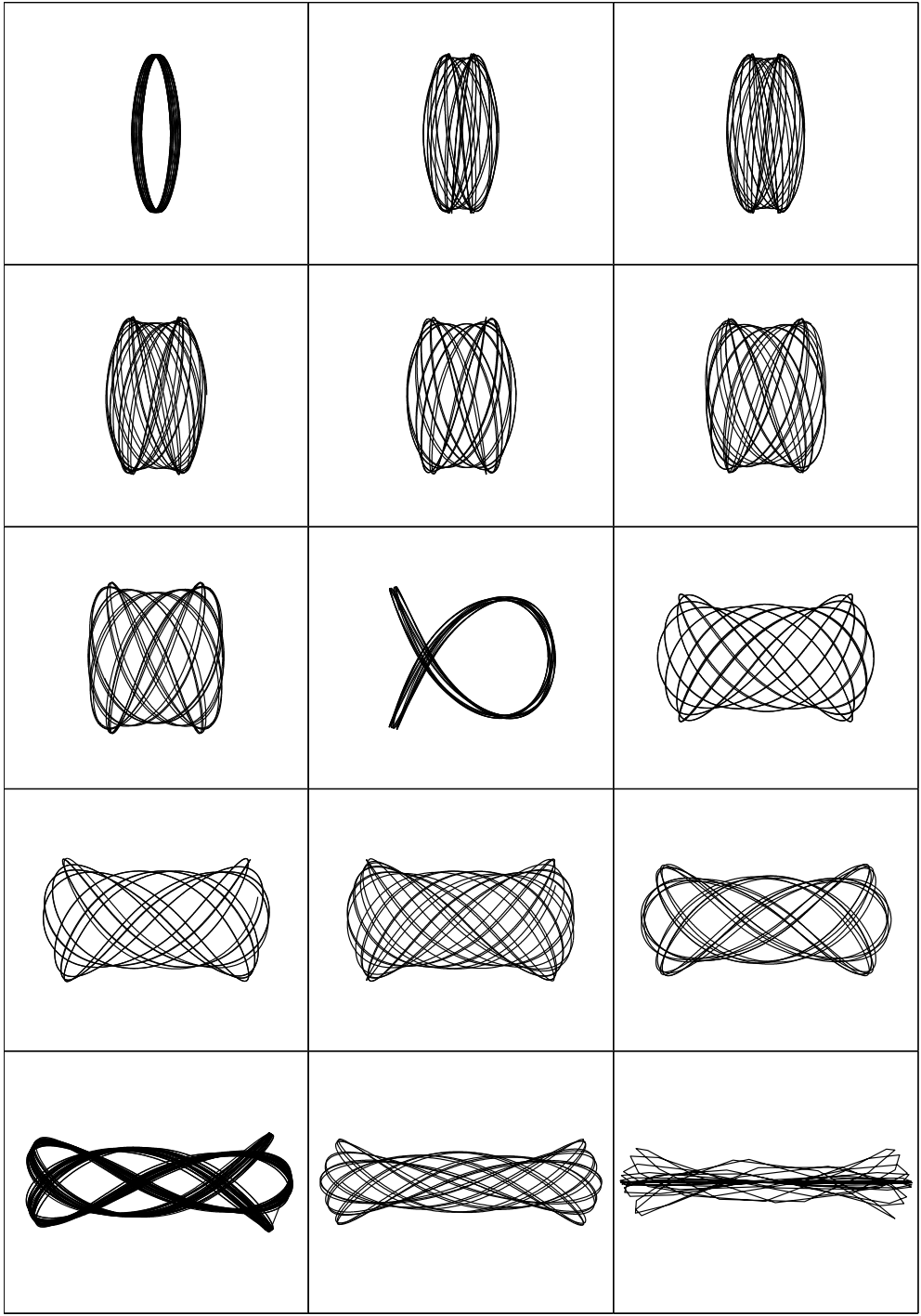} }} 
\figcaption{Collection of orbits for triaxial potential with axis
parameters $(a,c) = (\sqrt{3},\sqrt{2}/2)$. The panels are labeled by
letters `a' -- `o', where the labels coincide with those on the 
surface of section plot shown in Figure 1c. }  
\label{fig:gallery} 
\end{figure}


To illustrate the variety of possible orbits, the third surface of
section in Figure \ref{fig:section1} includes labels for 15 points in
the phase plane (indicated by the letters `a' -- `o'). Each of these
points is associated with a corresponding orbit, which is shown in the
15 panels of Figure \ref{fig:gallery}. Orbit `a' is located in the
central region of the phase plane, i.e., the region that dominates for
nearly spherical potentials (see Fig. \ref{fig:section1}); the
corresponding orbit in Figure \ref{fig:gallery} is thus a tube
orbit. The tube orbits become increasingly complicated (see orbits `b'
and `c') as one moves out from the central region. The remaining parts
of the surface of section primarily correspond to box orbits of
varying degrees of complexity. For orbits that lie in the centers of
islands in the phase plane, the corresponding orbits show symmetry --
for example, orbit `i' is a `fish orbit'. Notice that orbits that lie
at the centers of islands in the phase plane exhibit this symmetry
because they represent ``resonant'' orbits. For example, the fish
orbit in Figure \ref{fig:gallery} corresponds to a 2:3 resonance in
the $x-z$ plane. Note that the orbits in Figure \ref{fig:gallery} were
constructed for purposes of illustration, so that ``tighter'' fish
orbits, closer to exact resonance, can be studied. One should also
keep in mind that these orbits are confined to the $x-z$ plane;
additional types of orbits can be obtained when the orbit is allowed
to explore the full three dimensions of the potential (BT87, PM01).
The instability of these orbits to motion in the perpendicular 
direction is considered in the following section. 

\begin{figure}
\figurenum{3}
{\centerline{\epsscale{0.90} \plotone{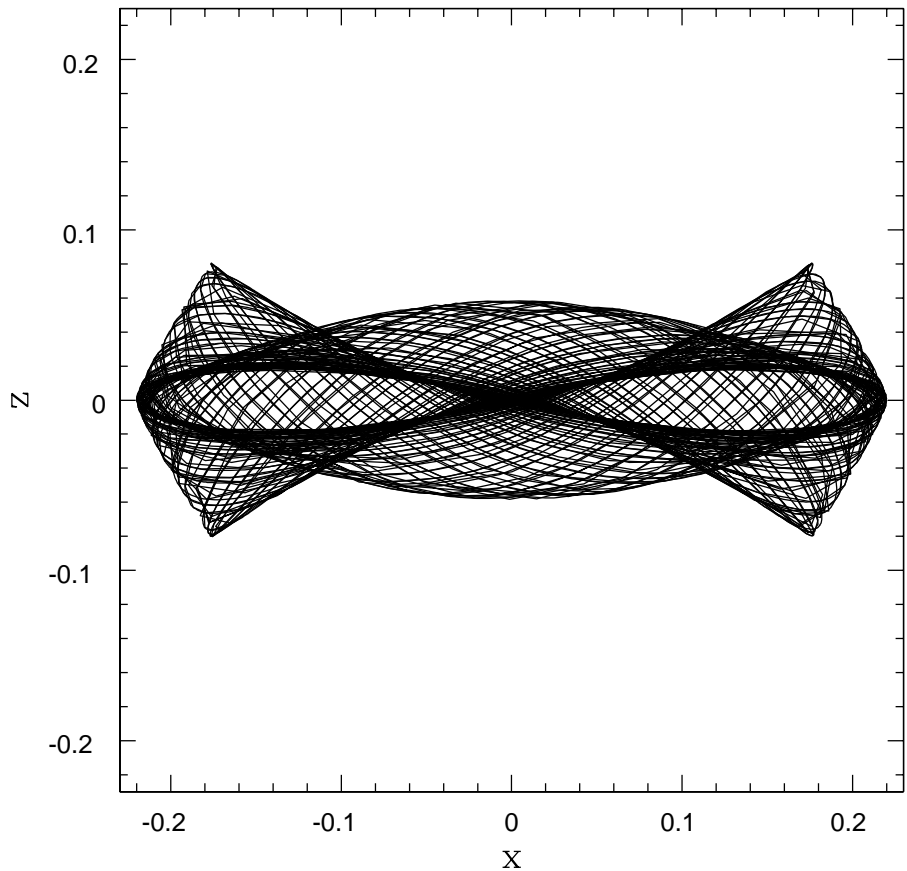} }}
\figcaption{Box orbit initially confined to the $x-z$ plane.  The
potential is taken to have axes $(a,b,c) = (\sqrt{2}, 1, 0.5071)$.
For this orbit, the time average $\langle j \rangle$ of the specific
angular momentum approaches zero, so the orbit has no sense of
rotation and is thus a box.  The orbit crosses near the origin in a
quasi-periodic manner, which leads to quasi-periodic forcing of the
perturbations in the perpendicular direction (see text). }
\label{fig:boxorbit} 
\end{figure}

Figure \ref{fig:boxorbit} shows a typical box orbit in this triaxial
potential. The axis ratios are chosen to be moderately triaxial, but
not extremely so.  This orbits displays a combination of regularity
and chaotic behavior. The orbit crosses back and forth across the
origin with no well-defined sense of rotation. The crossing time scale
is of order unity (in dimensionless units), although it varies
slightly from cycle to cycle. The distance of closest approach to the
origin also changes from cycle to cycle, but exhibits greater
variation.  In the discussion of orbital instability that follows, we
use this ``typical'' orbit as a benchmark case to study the
instability.

\section{INSTABILITY} 

\begin{figure}
\figurenum{4}
{\centerline{\epsscale{0.90} \plotone{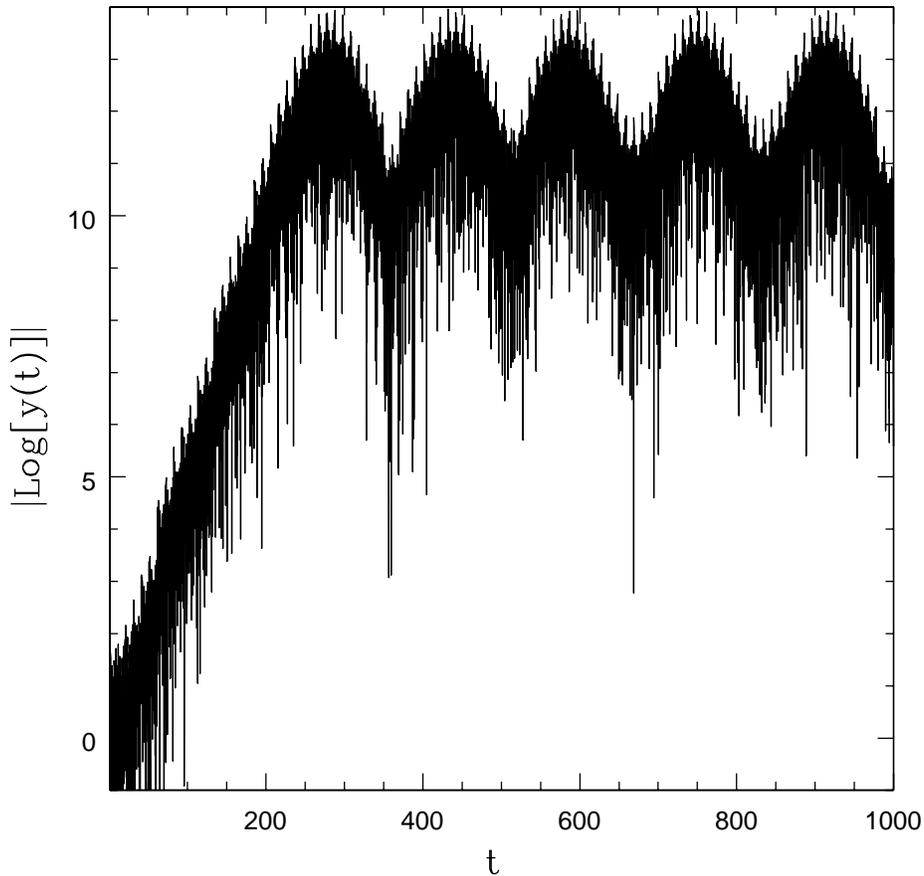} }}
\figcaption{ Development of the instability for the box orbit shown in
Figure \ref{fig:boxorbit}. The original orbit lies in the $x-z$ plane
with a small perturbation in the perpendicular direction ($\Delta y
\sim 10^{-8}$). The amplitude of the perpendicular coordinate (the
$\hat y$ coordinate) thus starts at $|y| \sim 10^{-8}$ and grows to
saturation during $\sim100$ time units. Note that we take the absolute
value of $y(t)$ and that we scale the coordinate by its initial value, 
so the vertical axis represents the growth factor.  Notice also that
the time evolution of the perpendicular coordinate simultaneously 
displays nearly periodic behavior, exponential growth, and chaotic
variations; at late times the exponential growth saturates and is
replaced by long-period oscillatory behavior. } 
\label{fig:unstable}
\end{figure}

A numerical survey of parameter space reveals that many of the orbits
that are initially confined to one of the principal planes are
unstable to motions in the perpendicular direction. One example of
this unstable behavior is shown in Figure \ref{fig:unstable}. In this
case, the orbit begins in the $x-z$ plane, but has a small
perturbation in the $y$-direction (here $\Delta y = 10^{-8}$). As
shown in Figure \ref{fig:unstable}, the subsequent evolution of the
$y$ coordinate displays four types of behavior: [1] The most striking
property of the solution is the overall growth of the amplitude, i.e.,
the envelope shows exponential growth.  [2] The solution exhibits a
well-defined periodicity superimposed on the exponential growth.  This
periodicity is a reflection of the underlying (near) periodicity of
the original box orbit in the $x-z$ plane and is thus not unexpected.
However, notice that the periodicity has multiple frequencies. [3] The
solution displays chaotic irregularities superimposed on the otherwise
smooth (regular) function. [4] The growth saturates at a well-defined
time scale, and the solution subsequently undergoes long period
oscillations (with additional short period oscillations and chaotic
irregularities superimposed). In the analysis and discussion given
below, our goal is to elucidate the fundamental mechanism(s)
responsible for these four properties.

For orbits that begin in a principal plane, the equation of motion in
the perpendicular direction (out of plane) determines whether or not
the orbit is stable.  For a given plane, we can consider the
perpendicular coordinate value to be small, at least at the beginning
of the evolution.  For example, if the original orbit lies in the
$y-z$ plane, then the $x$ coordinate is small so that $|x| \ll
y,z$. In this limit, the equation of motion for the non-planar
direction takes the form
\be 
{d^2 x \over dt^2} + \omega_x^2 x = 0 \qquad {\rm where} \qquad 
\omega_x^2 = { 4/a \over \sqrt{c^2 y^2 + b^2 z^2} + a \sqrt{y^2 + z^2} } \ , 
\label{eq:omegax} 
\ee 
Similarly, for the other two principal planes, the equations of 
motion for the perpendicular coordinate can be written 
\be 
{d^2 y \over dt^2} + \omega_y^2 y = 0 \qquad {\rm where} \qquad 
\omega_y^2 = { 4/b \over \sqrt{c^2 x^2 + a^2 z^2} + b \sqrt{x^2 + z^2} } \ , 
\label{eq:omegay} 
\ee 
and 
\be 
{d^2 z \over dt^2} + \omega_z^2 z = 0 \qquad {\rm where} \qquad 
\omega_z^2 = { 4/c \over \sqrt{b^2 x^2 + a^2 y^2} + c \sqrt{x^2 + y^2} } \ . 
\label{eq:omegaz} 
\ee

\begin{figure}
\figurenum{5}
{\centerline{\epsscale{0.90} \plotone{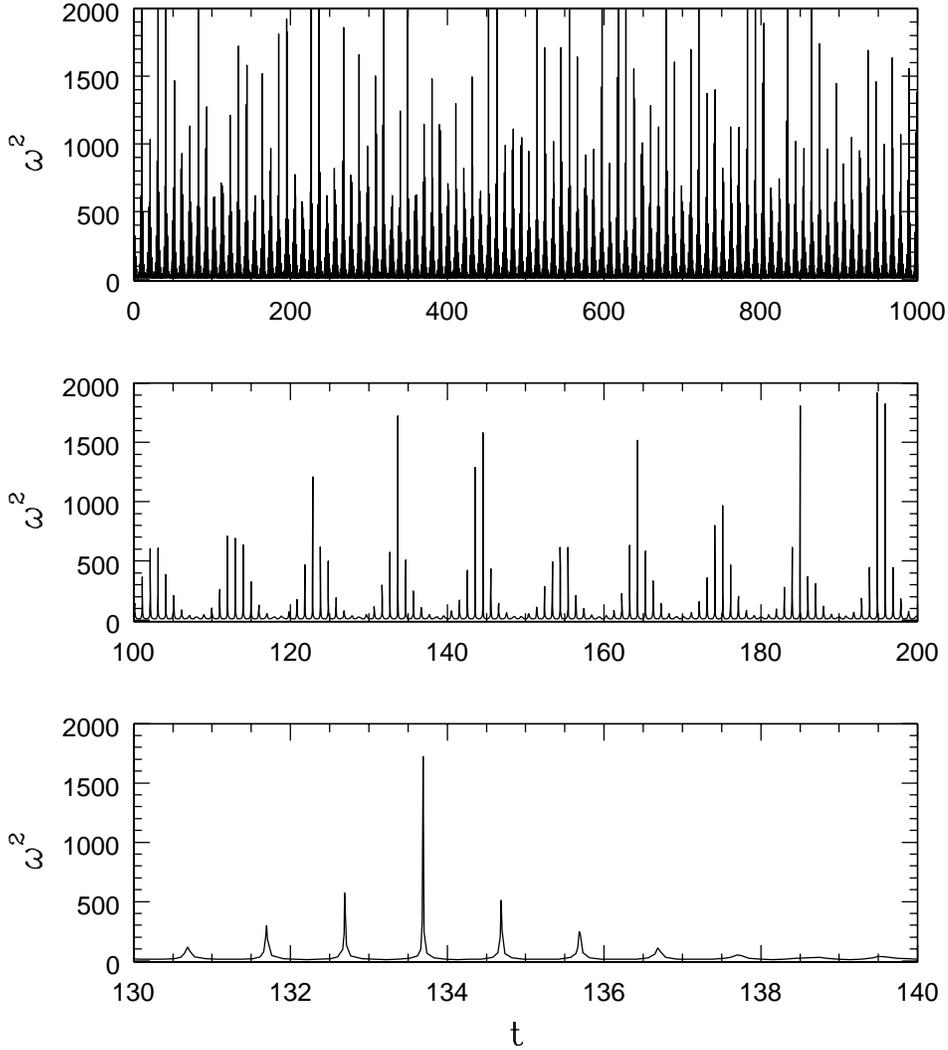} }}
\figcaption{ Driving frequency $\omega^2 (t)$ as a function of time
for the orbit depicted in Figures \ref{fig:boxorbit} and \ref{fig:unstable}. 
This plot represents the effective oscillation frequency of the test
particle in the direction perpendicular to the starting plane of the
orbit (here, the $\hat y$ direction). The three panels show three
temporal ranges, i.e., 1000 time units (top), 100 time units (middle),
and 10 time units (bottom).  Notice that the spikes in $\omega^2$ are
extremely narrow, much closer to the delta function limit than the
cosine limit. Notice also that the forcing function has structure on a
wide range of time scales, and shows both periodicity and chaotic
behavior. }  
\label{fig:omega} 
\end{figure}

\begin{figure}
\figurenum{6}
{\centerline{\epsscale{0.90} \plotone{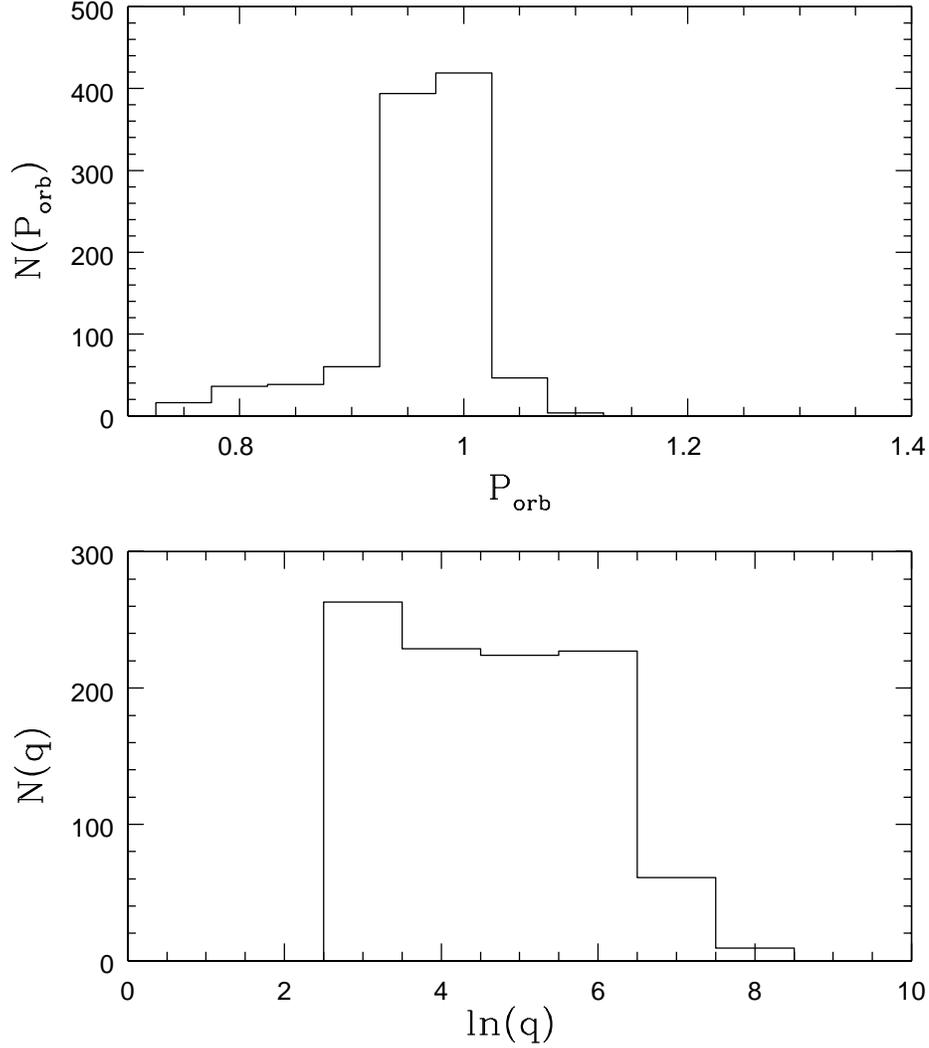} }}
\figcaption{ Histograms showing the distributions of periods (top
panel) and forcing strengths (bottom panel) for the orbit considered
in Figures 2 -- 4. The period is defined to be the time interval
between successive maxima in the function $\omega^2 (t)$ and the
forcing strength $q$ is defined to be the peak values of the function
(the maxima). Note that the scale for the period, and hence its
distribution, is narrow, with mean value $\langle P_{\rm orb}
\rangle$ = 0.986, and standard deviation $\sigma_P$ = 0.057. In
contrast, the distribution of forcing strength is shown using
logarithmic binning and spans a wide range, with mean value $\langle q
\rangle \approx$ 360 and standard deviation $\sigma_q \approx$ 560. }
\label{fig:qphist} 
\end{figure}
These equations of motion for the perpendicular coordinate
(eqs. [\ref{eq:omegax}--\ref{eq:omegaz}]) provide both a qualitative
and quantitative explanation for the instability, although the rest of
the orbit, described by the other coordinates [e.g., $x(t)$ and
$z(t)$], play a role. (In addition, in order to understand the
saturation mechansim, we must go beyond the limit of small
perpendicular coordinate values --- see below.)  To fix ideas, we
present the driving frequency $\omega_y^2$ for a box orbit in the
$x-z$ plane in Figure \ref{fig:omega}. This function $\omega^2 (t)$ is
typical for orbits that exhibit instability in the perpendicular
coordinate, although the exact form varies with the orbit under
consideration. Note that the driving frequency shows both periodic and
stochastic behavior.  Furthermore, this form for $\omega^2$ allows for
a qualitative description for the four properties of the unstable
solutions listed above:

[1] {\sl Instability.} At first glance, the equations look like those
of a simple harmonic oscillator, which would naively imply stability.
Upon closer inspection, however, the functions $\omega^2 (t)$ are
highly time variable and nearly periodic.  In the limit where the
functions $\omega^2 (t)$ are exactly periodic, the equations of motion
are a type of Hill equation (Hill 1886; Magnus \& Winkler 1966,
hereafter MW66; Abramowitz \& Stegun 1970, hereafter AS70). Since Hill
equations are known to have regimes of instability, the fact that we
see unstable solutions follows immediately. Notice that in the limit
of nearly circular orbits in a nearly spherical potential, the
equation of motion for the perpendicular coordinate becomes Mathieu's
equation (\S 5.3), whose regimes of instability are well studied
(AS70, Arfken \& Weber 2005, hereafter AW05). This Mathieu-equation
limit has been studied previously in a related context (Binney 1981,
Scuflaire 1995). Notice also that the driving frequency function
$\omega^2 (t)$ displays extremely narrow spikes, rather than smooth
cosine curves, as in Mathieu's equation, so the latter does not
represent a good approximation for most of the parameter space of
interest here. We thus consider the opposite limit in which the peaks
in the driving frequency function are considered as Dirac delta
functions (see \S 5; Appendices B and C).  

[2] {\sl Periodicity.}  The second key feature of the observed
unstable solutions is their periodicity. This feature also follows
directly from the fact that the perturbation equation is of the Hill
variety. In particular, Floquet's Theorem implies that the Hill
equation has solutions of the form
\be
y(t) = {\rm e}^{i \alpha t} p(t) \, , 
\ee 
where $p(t)$ is a periodic function and where the phase factor can
provide a growing (unstable) solution when $\alpha$ is imaginary
(MW66, AS70). These previous mathematical results imply that the
period of $p(t)$ is either the same as that of the driving frequency
$\omega^2 (t)$ or twice as great. In this context, the periodicity of
the driving frequency $\omega^2 (t)$ is determined by the period of the
orbit in the (original) orbit plane. This period, in turn, is given by
equations (\ref{eq:periodfull}) and (\ref{eq:periodin}) to a good
working approximation. As shown in Figure \ref{fig:omega}, and
indirectly by Figure \ref{fig:boxorbit}, the driving frequency has 
multiple periods; this complication leads to the solutions $y(t)$
displaying multiple periodicities as well.

[3] {\sl Chaos.} The motion of the perpendicular coordinate is
ultimately driven by the (time dependent) coordinates of the starting
orbit (in its initial plane). The amplitude of the driving frequency
oscillations $\omega^2 (t)$ is essentially determined by the distance of
closest approach of the orbit to the origin (see eqs.
[\ref{eq:omegax} -- \ref{eq:omegaz}]) and this distance varies from
passage to passage. If the orbit is chaotic, as is often the case,
these distances of closest approaches will be given by some
distribution of values, and the corresponding driving frequency will
have a chaotic element.

One might worry that departures from exact periodicity would
invalidate previous mathematical results concerning instability. This
issue led us to analyze the quasi-periodic Hill's equation with random
forcing strengths (Adams \& Bloch 2007, hereafter AB07). These new
results show that the departures from strict periodicity in Hill's
equation result in a re-scaling of the parameters, but do not
otherwise compromise the instability (see Theorem 1 of AB07; see also
Appendix D).  In addition, the stochastic variations in the forcing
function lead to two separate contributions to the growth rate: an
asymptotic growth rate that results from a direct application of
Floquet's theorem, and an anomalous growth rate that results from
matching the solutions from cycle to cycle, where each cycle has a
different forcing strength (see Theorem 2 of AB07).

The distributions of periods and forcing strengths thus play an
important role in determining the dynamics. For the box orbit depicted
in Figure \ref{fig:boxorbit}, and the corresponding forcing function
$\omega^2 (t)$ shown in Figure \ref{fig:omega}, we show the
distributions of periods and forcing strengths in Figure
\ref{fig:qphist}. For the sake of definiteness, we have defined the
period to be the time interval between peaks of the function
$\omega^2(t)$. The forcing strengths ($q$) are defined to be the
heights of the peaks. Although the distributions of period and forcing
strength vary with the orbit under consideration, Figure
\ref{fig:qphist} reveals several important and typical trends: The
distribution of periods is quite narrow, so that departures from
strict periodicity are small. This finding makes the corrections for
varying periods corrrespondingly small (where this statement is
quantified in Appendix D). On the other hand, the distribution of
forcing strengths $q$ is wide, with the standard deviation larger than
the mean. Both the mean value (appropriately weighted) and the width
of the distribution play a role in determining the growth rates for
instability.

\begin{figure}
\figurenum{7}
{\centerline{\epsscale{0.90} \plotone{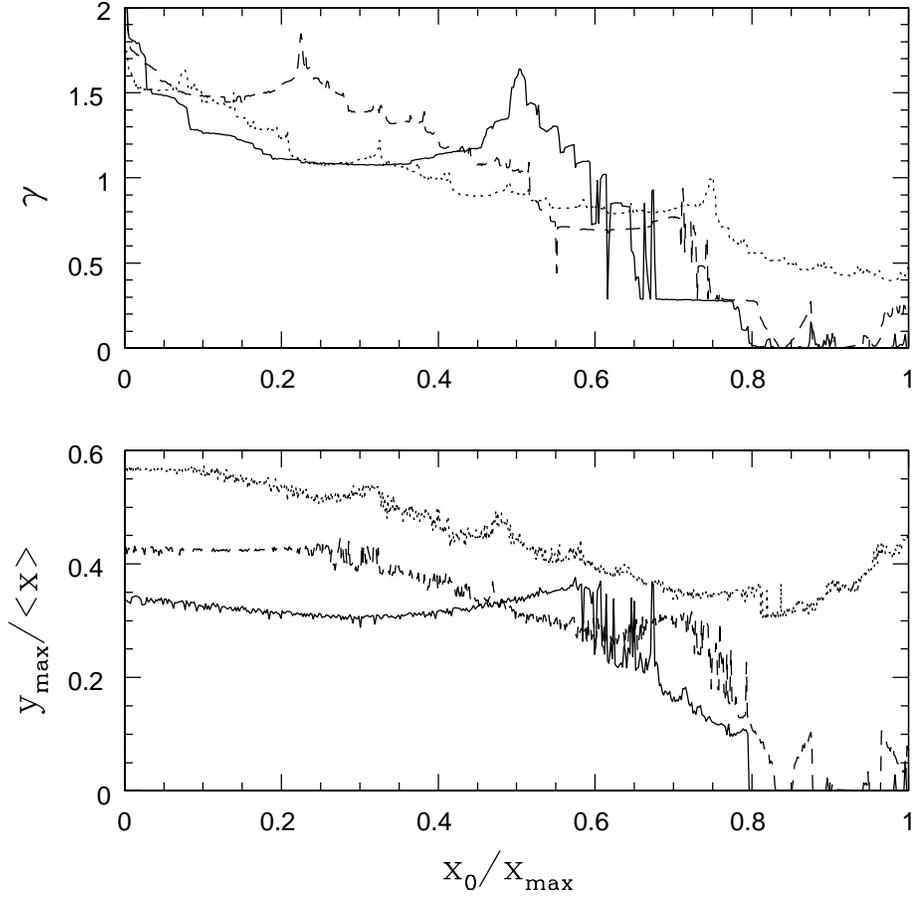} }}
\figcaption{ The upper panel shows the growth rate of instability as a
function of starting coordinate. For all of the cases considered here,
the energy is taken to be $E$ = 0.1, and the axis variables $b$ = 1
and $c = \sqrt{2}/2$. Solid curve shows the growth rate $\gamma$ for
varying starting locations for the axis parameter $a$ = $\sqrt{3/2}$;
the dashed and dotted curves show the results for $a$ = $\sqrt{3}$ and
$a$ = $\sqrt{6}$, respectively. The bottom panel shows the
corresponding saturation level, i.e., the maximum value of the $y$
coordinate expressed as a fraction of the typical $x$ coordinate. The
three curves correspond to the three values of $a$ in the top panel. }  
\label{fig:gamvsx} 
\end{figure}

In addition to chaotic behavior within a given orbit, the variation of
dynamical properties varies from orbit to orbit, even those with
almost the same starting conditions. The orbits thus display sensitive
dependence on their initial conditions, one of the defining properties
of chaotic systems. As one example of this behavior, Figure
\ref{fig:gamvsx} shows the growth rate of the instability plotted as a
function of the starting $x$-coordinate. This plot shows that nearby
starting trajectories can result in significantly different growth
rates. Since the growth of the $y$-coordinate is not perfectly
exponential, some ambiguity arises in the definition of the growth
rate. The results shown in Figure \ref{fig:gamvsx} were obtained by
finding the maximum value $|y|_{max}$ of the $y$-coordinate over the
first 50 time units, and defining $\gamma \equiv \ln |y_{max}| /50$;
these growth rates thus represent the average growth rates over this
time interval.  Figure \ref{fig:gamvsx} also shows the saturation
level of the perpendicular coordinate (in the lower panel) plotted as
a function of the same starting conditions. Here, the saturation
levels were computed by finding the maximum displacement over the
first 200 time units (thereby allowing the instability time to
saturate).  The behavior of the saturation level is also highly
structured, i.e., not a smooth function of $x_0/x_{max}$. The fine
scale structures in both the growth rate and the saturation level
(when considered as a function of $x_0/x_{max}$) arise (in part) from
the narrow bands of stability that are always present in the Hill's
equation that describes the instability (see AS70, Weinstein \& Keller
1987, AB07, and \S 5 below).  Further, we find that the saturation
levels correlate with the growth rates; in particular, low or
vanishing growth rates correspond to low saturation levels, as
expected.  Figure \ref{fig:gamvsx} shows additional systematic
behavior, in that the ranges of starting conditions that lead to large
growth rates and large saturation levels are relatively well-defined.
The larger structures in the growth rate functions (and saturation
levels) often coincide with the structure of the phase-plane diagram
for this potential.
[4] {\sl Saturation.} The strength of the instability, as well as its
existence, depends on the amplitude of the driving frequency
$\omega^2 (t)$.  The equations of motion found above are derived in the
limit where the perpendicular coordinate is small compared to the
coordinates in the (initial) orbit plane. As the perpendicular
coordinate grows, this approximation becomes invalid, and the relevant
equation of motion takes on a modified form. For the cases of interest
here, the driving frequencies have the basic form $\omega^2 (t) \sim
1/\xihat$, where $\xihat$ is a modified ``radial coordinate'' where
the $(x,y,z)$ coordinates are weighted differently, by the geometric
factors ($a,b,c$).  As the perpendicular coordinate (e.g,. $y$) grows,
its contribution to $\xihat$ and hence the driving function becomes
significant, with the net effect being to reduce the forcing
amplitude. When the amplitude decreases enough, the motion is no
longer unstable in the perpendicular coordinate, and the instability
saturates.

As a particular example of the behavior described above, we consider
the equation of motion for the $\hat y$ coordinate for an orbit
initially confined to the $x-z$ plane in the limit of a nearly
spherical potential (eq. [\ref{eq:phidelt}]), i.e., 
\be
{d^2 y \over dt^2} + {1 \over 2 \xi} \Bigl[ 1 + 
{\delta (x^2 - z^2) \over 4 \xi^2} \, \Bigr] y = 0 \, .  
\ee 
When the $y$ coordinate is small (compared to $x$ and $z$), the size
of the driving term is determined by $1/\xi \sim (x^2 + z^2)^{-1/2}$.
The forcing term is large for small values of displacement $\xi$. 
When the $y$ coordinate grows, however, its contribution to $\xi$ 
becomes significant (i.e., $\xi$ is not necessarily small even when 
$|x|, |z| \ll 1$), and the instability saturates. 

To see this trend explicitly, let $\xi = (d^2 + t^4)^{1/2}$, where
this form assumes that the orbit has a distance of closest approach
$d$ and the radius of the orbit is given by a constant force law
(appropriate for radial orbits -- see \S 3.2). For a nearly linear
orbit with closest approach $d$, the radial distance is given by
$\xi^2 = d^2 + s^2$, where $s$ is the distance along the orbit
measured from the point of closest approach.  Since $s \propto t^2$
for a constant force law, one obtains the relation $\xi^2 = d^2 + t^4$
in appropriate dimensionless units.  The effective amplitude of the
driving frequency is then given by the integral
\be 
q = \int_{-\infty}^{\infty} {dt \over (d^2 + t^4)^{1/2} } = 
{8 \, \Gamma^2(5/4) \over \sqrt{\pi d \,} }
\approx {3.71 \over \sqrt{d \,} } \, . 
\ee 
At the start of the evolution, when the motion is confined to a plane,
the distance of closest approach to the origin $d = d_0$ is given by
that of the starting orbit. For a box orbit, $d_0$ can be quite small,
so that $q$ is large. A general property of Hill's equations is that
larger $q$ implies greater instability (although islands of stability
will remain -- see MW66, AS70, and \S 5 below). Once the instability
has developed, the distance of closet approach must include the
displacement in the perpendicular direction, so that $d \to \ybar$
(where the angular brackets represent an appropriate average of the
perpendicular coordinate).  The value of $q$ at later times is thus
reduced by a factor $(d_0 / \ybar)^{1/2}$. Notice that saturation of
the instability requires that $\ybar \gg d_0$, which is much less
restrictive than the requirement $\ybar \sim d_{max}$. In other words,
the instability can saturate when the perpendicular displacement,
measured here by $\ybar$, becomes sufficiently larger than the
(initial) distance of closest approach, but need not be comparable to
the full orbit size. In many cases, however, the saturation levels 
are large enough that $\ybar$ is comparable to the original orbital 
extent in the $x-z$ plane (see Fig. \ref{fig:gamvsx}). 
\begin{figure}
\figurenum{8a}
{\centerline{\epsscale{0.90} \plotone{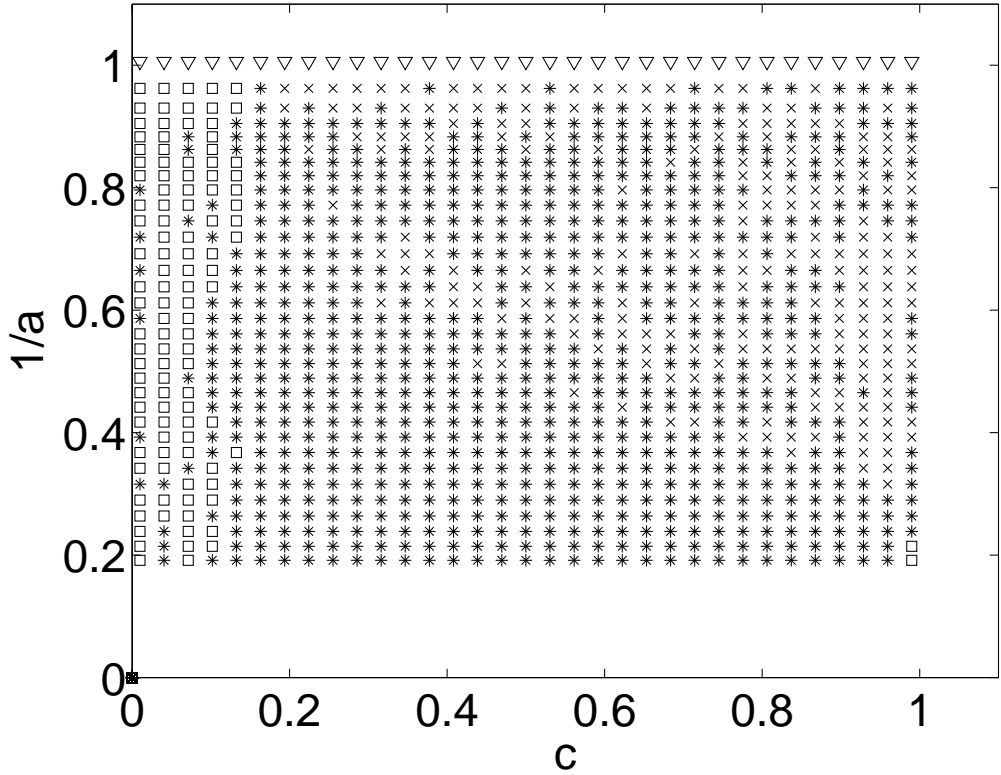} }} 
\figcaption{Instability as a function of axis weights. Nearly
spherical potentials correspond to the upper right corner of each 
panel.  For each point in the grid, orbits are started in the $x-z$
plane, with fixed energy, and with a fixed starting value of the $\hat
x$ coordinate. The symbols denote the type of resulting orbits, where
open squares are stable box orbits, open triangles are stable loop
orbits, stars are unstable box orbits, and crosses are unstable loop
orbits.  (a) Result for $x_0/x_{max}$ = 0.10.  (b) Result for
$x_0/x_{max}$ = 0.50.  (c) Result for $x_0/x_{max}$ = 0.90. } 
\label{fig:acplane01} 
\end{figure}

{\centerline{\epsscale{0.90} \plotone{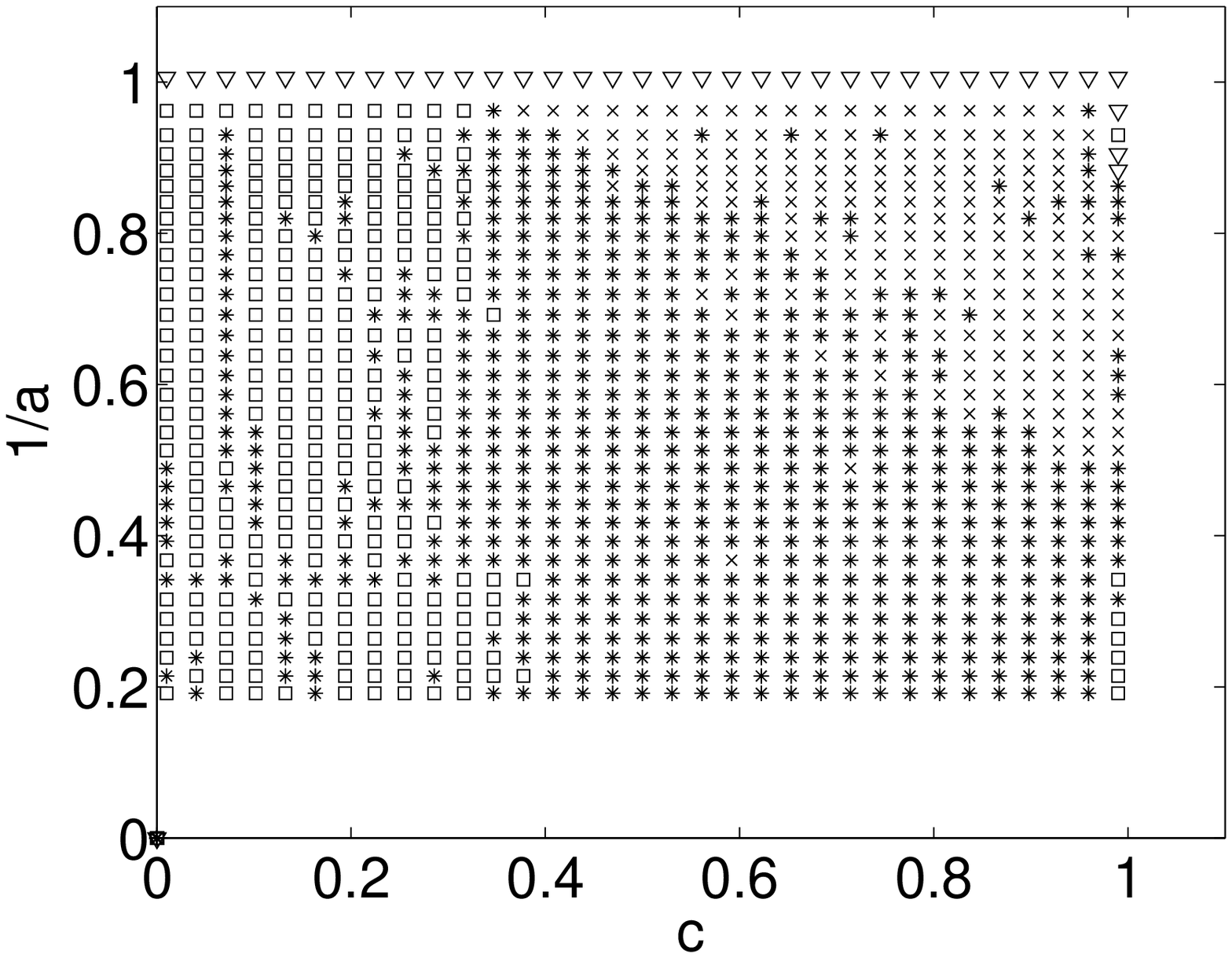} }} 
\centerline{Fig. 8 --- Continued.} 


{\centerline{\epsscale{0.90} \plotone{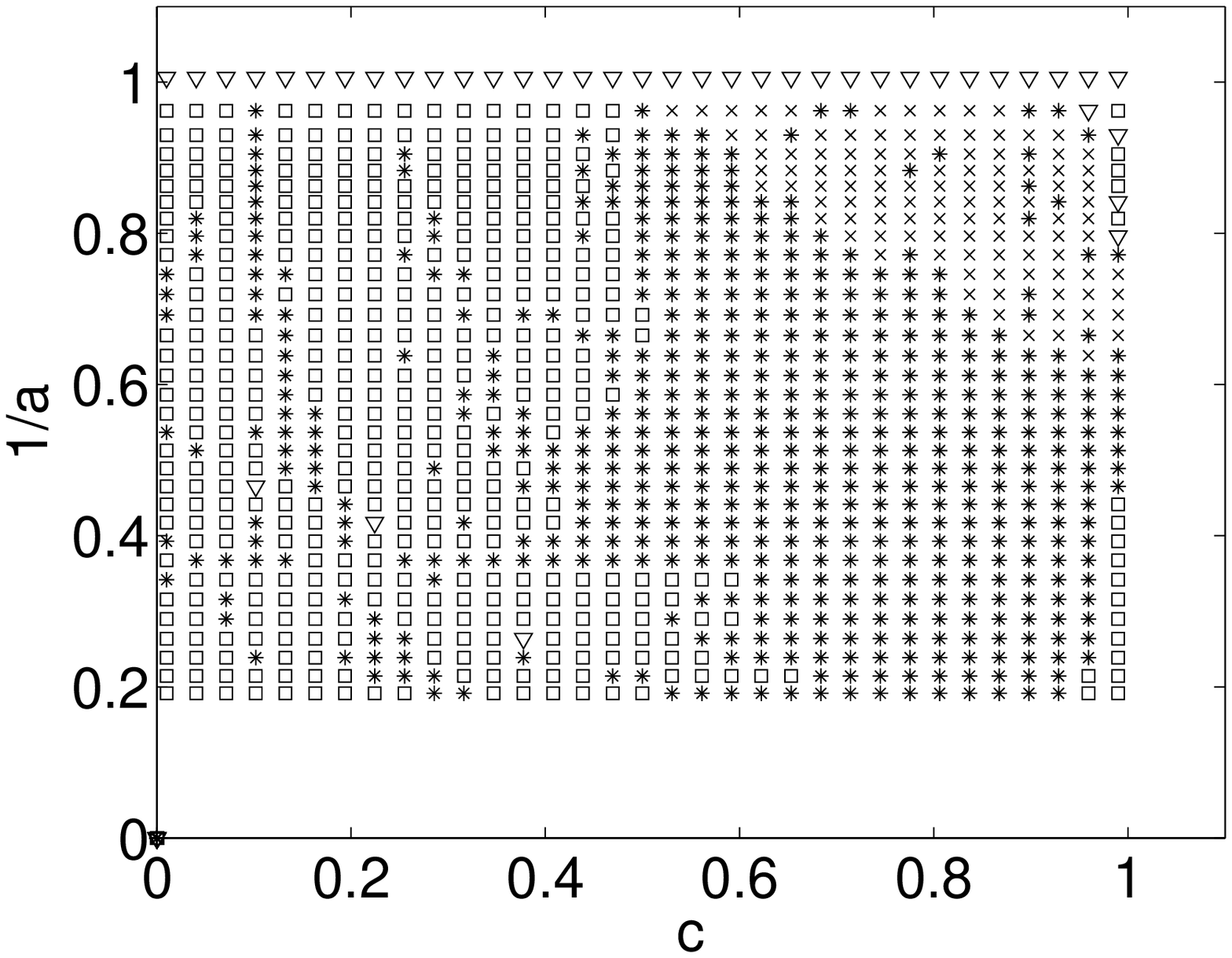} }} 
\centerline{Fig. 8 --- Continued.} 


Before leaving this section we consider the degree of stability versus
instability as a function of the shape of the triaxial potential,
where shape is codified by the axis weights $(a,b,c)$. For a given
energy and set of axis weights, we start orbits with a given value of
$x_0/x_{max}$ as described in \S 3. We then determine if the orbit is
a loop orbit or a box orbit, and whether the orbit is stable or not.
To determine the orbit class, we use a procedure similar to that
developed by Fulton \& Barnes (2001).  For a given orbit, we fourier
analyze both time series $x(t)$ and $z(t)$ and obtain the dominant
frequencies.  If the dominant frequencies are in 1-1 correspondence,
the orbit is labeled as a loop (where we allow for a 5\% discrepancy
in the frequency ratio). To determine if the orbit is unstable, we
monitor the maximum displacement $|y|_M$ in the perpendicular
direction. If the perpendicular coordinate grows above a given
threshold, then the orbit is considered unstable. Since orbits
generally either grow and saturate at large values of $|y|_M$, or
remain stable with small values of $y \sim 10^{-8}$, the results are
insensitive to the threshold value. For the sake of definiteness we
use a threshold $|y|_M/r_M \ge 0.01$ (where $r_M$ is the maximum
two-dimensional radius in the original $x-z$ plane of the orbit). The
results are shown in Figure \ref{fig:acplane01} for three different
values of the starting $\hat x$ coordinate.  Several trends are
evident. In all cases, a significant fraction of the plane shows
unstable orbits, so that the instability considered herein is indeed
robust.  As the starting value $x_0/x_{max}$ is increased, the orbits
become more nearly ``radial'' (following the $\hat x$ axis) and the
degree of instability decreases somewhat (compare the panels of Fig.
\ref{fig:acplane01}); further, the regimes of stability (usually)
correspond to extreme axis weights (small $c$ and large $a$). Finally,
we note that the regimes of stability and instability are intermixed,
as expected in highly chaotic systems.

\section{MODEL EQUATIONS} 

In this section we develop and analyze a class of model equations that
captures the essential physics of the orbit instability found above.
As shown in the previous section, when the motion is initially
confined to one of the principal planes, the equation of motion for
the amplitude of the perpendicular coordinate takes the form
\be 
{d^2 y \over dt^2} + [ \af + Q(t) ] y = 0 \, ,
\label{eq:modeleq} 
\ee 
where $Q(t)$ is nearly periodic. Here we use the variable $y$ as the
perpendicular coordinate, but the same general form holds for the
perturbation direction falling along any of the axes. Note that we
have separated out part of the forcing function as a baseline
frequency $\sqrt{\af}$, although this decomposition is not unique.
Notice also that the function $Q(t)$ can be considered as a function
of the form $Q[x(t), z(t)]$, where $x(t)$ and $z(t)$ describe the
orbital motion in the original plane, so that $Q$ is given by
equations (\ref{eq:xforce}--\ref{eq:zforce}) in general and by
equations (\ref{eq:omegax}--\ref{eq:omegaz}) in the limit of small 
$y$ (as assumed here). 

In most applications, $Q(t)$ will be nearly periodic because its time
variation is given by an orbit solution, although the crossing time of
the orbit can vary from cycle to cycle.  In this context, in the
Hernquist potential, the orbital periods depend on energy, but are
nearly independent of angular momentum (see Fig. 3 of AB05),
especially in the inner limit as considered here, so that the cyclic
forcings have nearly the same time intervals. This expectation is
borne out in numerical evaluations of the orbits, as shown in the
distributions of Figure \ref{fig:qphist}.  On the other hand, the
amplitude (the forcing strength) varies substantially from cycle to
cycle. If the function $Q(t)$ is exactly periodic, then the relevant
equation of motion has the form of Hill's equation (Hill 1886), and
we can draw on a number of well-known results (MW66, AS70, Arnold
1978). In this section we consider the periodic version of Hill's
equation as a first approximation, and then show how to generalize the
results to cases where the forcing strength varies from cycle to
cycle.  This section is augmented by Appendices B -- D, which include
some of the mathematical details, and by Appendix E, which presents a
heuristic review of Hill's equation.  We consider a more rigorous
mathematical treatment of the stochastic problem in a separate
publication (AB07).

\subsection{The Delta Function Limit} 

As a first approximation, we consider the forcing potential to be
sufficiently sharp that we can model $Q(t)$ with a Dirac delta
function. In physical terms, this approximation means that the orbit
passes close to the origin (so that the spike in $Q(t)$ is large) and
that the orbit spends little time near the origin (so the spike is
narrow). Both of these conditions are met for box orbits in triaxial
potentials. In this limit, the equation of motion for the
perpendicular coordinate takes the particular form 
\be 
{d^2 y \over dt^2} + [ \af + q \delta (\tmod - \pi/2) ] y = 0 \, ,
\label{eq:eqdelta} 
\ee 
where $q$ measures the strength of the spike and where $\delta$ is the
Dirac delta function. Here we have scaled the time variable so the
period of one cycle is $\pi$.  Note that the argument of the delta
function is written in terms of $\tmod$, which corresponds to the time
variable mod-$\pi$, so that the forcing potential is $\pi$-periodic.
Notice also that the actual physical problem has a forcing strength
$q$ that varies from cycle to cycle. As a result, we should think of
the ``constant'' $q$ as a stochastic variable that takes on a new
value every cycle $0 \le \tmod \le \pi$.

\begin{figure}
\figurenum{9}
{\centerline{\epsscale{0.90} \plotone{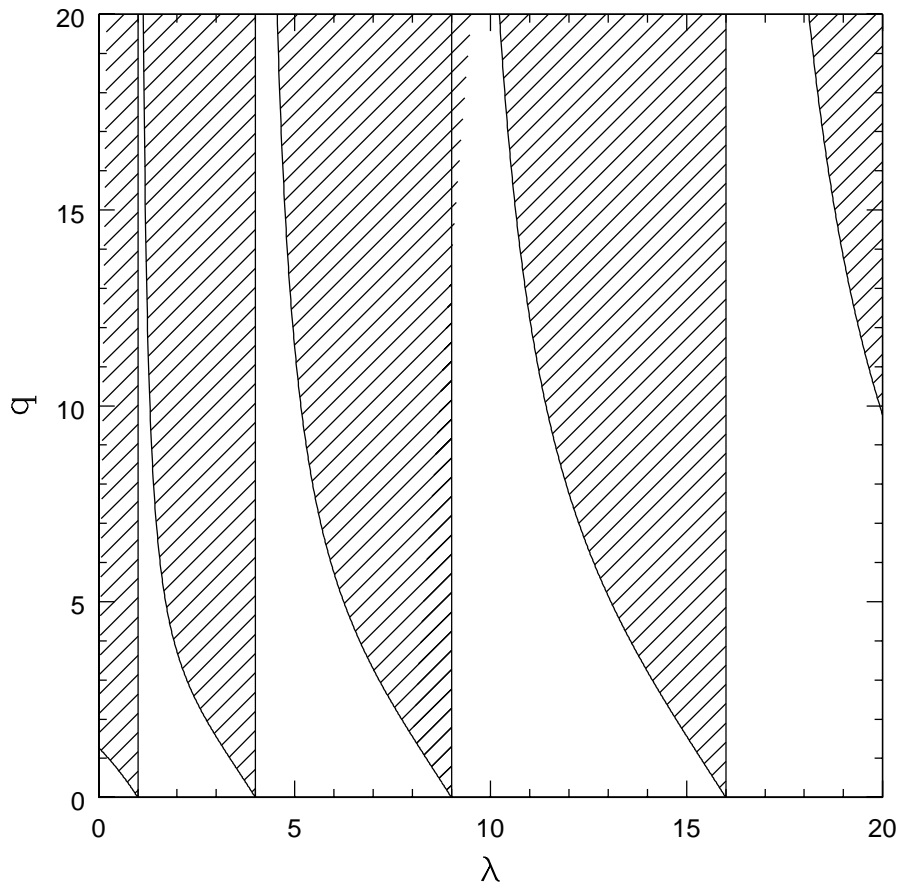} }} 
\figcaption{ Regions of instability for Hill's equation in the delta
function limit.  The shaded regions show the values of $(\lambda,q)$
that correspond to exponentially growing (unstable) solutions, which 
represent unstable growth of the perpendicular coordinate for orbits 
in our triaxial potential that are initially confined to one of the 
principal planes (see also AB07). } 
\label{fig:qlplane} 
\end{figure}

In the delta function limit, the solution to Hill's equation is thus
specified by two parameters: the frequency parameter $\lambda$ and the
forcing strength $q$. The solutions to equation (\ref{eq:eqdelta}) are
constructed in Appendix B along with the determination of the growth
rates for instability.  Figure \ref{fig:qlplane} shows the plane of
possible parameter space for Hill's equation in this limit, with the
unstable regions shaded. Notice that a large fraction of the plane is
unstable. This result is in keeping with our previous finding that the
equation of motion for the instability takes the form of a Hill's
equation with highly spiked forcing functions and that a large
fraction of the orbits are unstable (\S 4).

An important feature of these solutions is that in the limit of
large forcing strength $q \gg 1$, the bands of stability become
extremely narrow --- but they never vanish entirely.  Physical orbits
are often found in this limit, as illustrated by Figures
\ref{fig:omega} and \ref{fig:qphist}.  This property is generic to
Hill's equations in the unstable (here, large $q$) limit (Weinstein \&
Keller 1987). For the case of delta function forcing terms, for
example, the widths of the stable zones are given by $(\Delta \lambda)
= (8 n^2/\pi) \langle 1/q \rangle$, where $n$ labels the order of the
stability strip (AB07).  The presence of these narrow zones implies
that the issue of stability/instability can depend sensitively on the
orbit parameters, i.e., nearby orbits can display significantly
different behavior. This trend is seen in the numerical results for
the growth rates, as depicted in Figure \ref{fig:gamvsx}, where the
function $\gamma (x_0/x_{\rm max})$ displays a great deal of
structure, so that orbits with nearly identical starting conditions
can have significantly different growth rates.

With the solution to the model equation in hand, as illustrated by
Figure \ref{fig:qlplane}, it is useful to reconnect with the original
orbit problem. In terms of physical orbits, the period of the
(periodic) forcing potential is determined by the crossing time
$t_{cross}$ of the orbit. Here we have defined the dimensionless time
variable $t$ so that the period is $\pi$; the physical time variable
$t_{phys}$ is thus related to the dimensionless time appearing in
equation (\ref{eq:eqdelta}) via $t_{phys} = t (t_{cross}/\pi)$.
Further, the period of a half orbit for the spherical limit is given
(almost) analytically in AB05,
\be 
\tau_{1/2} = (4 \pi G \rho_0)^{-1/2} \epsilon^{-3/2} 
\left\{ \cos \sqrt{\epsilon} + \sqrt{\epsilon} 
\sqrt{1 - \epsilon} \right\} {\cal A} \, , 
\ee 
where $\epsilon = |E/\Psi_0|$ is the dimensionless energy and where
the correction factor $\cal A$ lies in the range $1 \le {\cal A} \le
1.05$ (see Fig. 3 of AB05). The strength $q$ of the forcing potential
is determined by the distance of closest approach of the orbit to the
origin ($q \sim 1/d$). In the spherical limit, the distance of closest
approach is given by the inner turning point of the orbit and is a
simple function of energy and angular momentum (AB05). Here, for loop
orbits, the orbit has an analogous inner turning point and the
spherical results can be used as a good starting approximation.  For
box orbits, the test particle wanders arbitrarily close to the origin,
so that the distance of closest approach will often be small (and
hence the parameter $q$ will be large); in general, $q$ can vary from
cycle to cycle. The problem of Hill's equation with random variations
in forcing strength is addressed in Appendix C for the delta function
limit (see AB07 for a more general treatment).

\subsection{Beyond the Delta Function Approximation} 

The physical orbits do not produce perfect delta functions for the
forcing potential $Q(t)$. Instead, the spikes have a small but finite
width. In order to understand how this feature affects the growth
rates (in particular, the regions of stability and instability), we
model the forcing function through an equation of the form
\be
{d^2 \xi \over d t^2} + \bigl[ \af + q_n \sin^{n} t \bigr] \xi = 0 \, , 
\label{eq:hilln} 
\ee
where $n$ is a (generally large) even integer. For the class of orbits
considered in the previous section, we have fit functions of the form
$\sin^n t$ to the functions $\omega^2(t)$ resulting from numerical
integration of the equations of motion. The values of $n$ required to
fit these results fall in the range $n = 10 - 100$, with a ``typical''
best-fit value $n \approx 64$. The full width at half maximum (FWHM)
of the spike is given by $w \approx 2 ([2 \ln 2]/n)^{1/2}$ in the limit
of large $n$ where the spikes have narrow widths.  For $n$ = 64, for
example, we obtain $w \approx 0.294$.

To compare model equations of this form for different values of $n$,
or to compare with the delta function limit, one needs to define the
effective forcing strength for a given value of $q_n$.  In the delta
function limit, this term has the form $q \delta([t]-\pi/2)$. Since
the delta function integrates to unity over one cycle, we can define 
the effective strength $q_{\rm eff}$ for a given $n$ through the relation
\be
q_{\rm eff} = q_n \int_0^\pi \sin^n dt = q_n \pi {(n-1)!! \over n!!} \, . 
\label{eq:renorm} 
\ee 

\begin{figure}
\figurenum{10}
{\centerline{\epsscale{0.90} \plotone{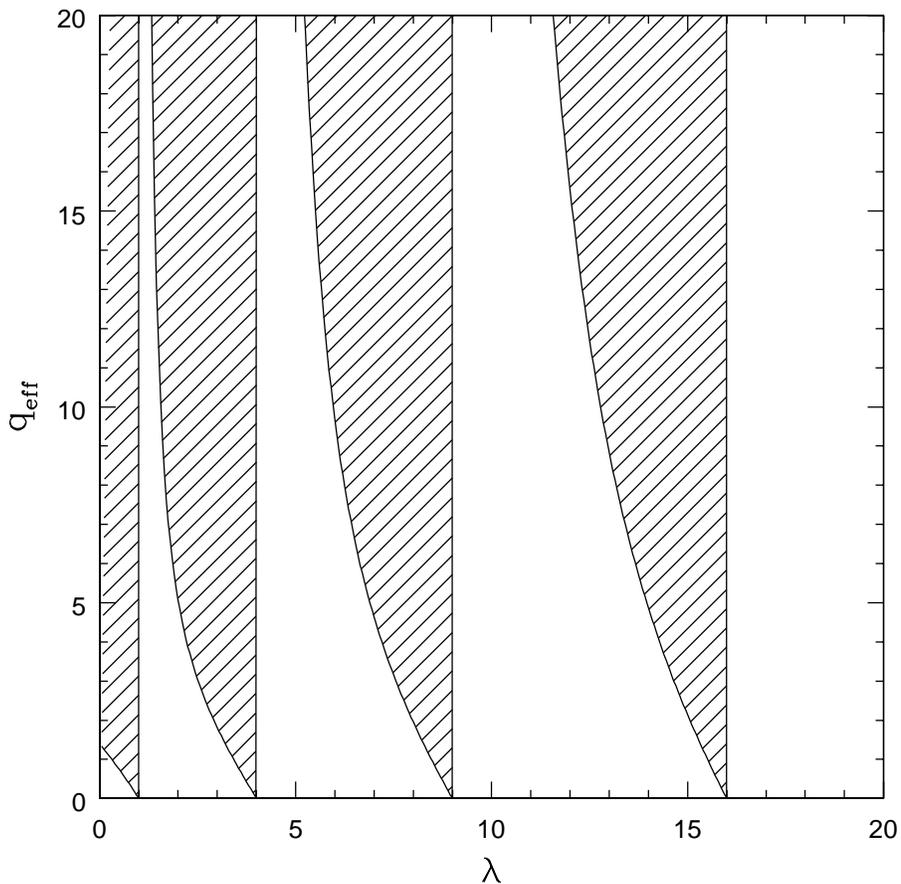} }} 
\figcaption{ Regions of instability for Hill's equation for periodic
spikes with nonzero width. The width is given by $\sin^n t$ where $n$
= 64.  The shaded regions show the values of $(\lambda,q_{eff})$ that
correspond to exponentially growing (unstable) solutions (compare
with Fig. \ref{fig:qlplane}). Note that the values of $q_{eff}$ have 
been normalized according to equation (\ref{eq:renorm}). } 
\label{fig:qlplaneb}
\end{figure}

Figure \ref{fig:qlplaneb} shows the plane of parameters $(\lambda,
q_{eff})$ for a Hill's equation of the form given by equation
(\ref{eq:hilln}) with $n$ = 64. The $q$ values have been scaled
according to the normalization of equation (\ref{eq:renorm}) so that
this plane can be directly compared with Figure \ref{fig:qlplane}.
After including this normalization, the two planes are more similar
than different, so that we expect the delta function model of the
previous subsection to provide a good description of the instability
mechanism. Nonetheless, Figures \ref{fig:qlplane} and
\ref{fig:qlplaneb} show slight differences, with the ``tongues of
instability'' being somewhat narrower for the case with $\omega^2$
spikes of finite width.


\subsection{The Nearly Spherical (Mathieu Equation) Limit} 

We note that previous treatments of similar orbit instabilities (e.g.,
Binney 1981, Scuflaire 1995) often find that the equation of motion
for the perpendicular coordinate (whose motion can be unstable) takes
the form of Mathieu's equation
\be
{d^2 y \over dt^2} +  (a + q \cos t) y = 0 \, . 
\ee 
Although this equation has qualitatively similar instabilities to
those considered here (compare the results presented in AS70 with
Figures \ref{fig:qlplane} and \ref{fig:qlplaneb}; see also Binney
1981), the forcing function $\omega^2 \sim \cos t$ is extremely
different from that shown in Figure \ref{fig:omega}. It is thus useful
to find the limiting regime of parameter space for which the
instability takes the form of Mathieu's equation, rather than the more
extreme, nearly delta function limit considered above.

In qualitative terms, in the limit of a nearly spherical system 
(\S 2.5), the equation of motion for the perturbation takes the 
simpler form 
\be
{d^2 y \over dt^2} + {2 \over \xi} y = 0 \, , 
\ee 
where $\xi^2 = x^2 + y^2 + z^2$ and where we have assumed 
motion initially confined to the $x-z$ plane. If we also 
consider the initial orbit to be nearly circular, then we 
can define an effective radial coordinate of the orbit 
$r^2 = x^2 + z^2$. The forcing term thus takes the form 
\be
\omega^2 = {2 \over \xi} \approx {1 \over r(t)} \approx 
{1 \over r_0 + (\Delta r) \cos \Omega t} \, , 
\ee 
where $r_0$ plays the role of the semimajor axis of the orbit, the
turning points of the orbit fall at $r_0 \pm \Delta r$, and $\Omega$
is the mean motion of the orbit (since the orbit is not an ellipse,
these quantities have slightly different meaning than in the standard
Kepler problem -- see AB05 for further discussion). With one further 
approximation, assuming $\Delta r \ll r_0$, the equation of motion 
for the instability takes the form 
\be
{d^2 y \over dt^2 } + {2 \over r_0} 
\Bigl[ 1 - {\Delta r \over r_0} \cos \Omega t \Bigr] y = 0 \, , 
\ee
which has the form of Mathieu's equation.  Although this result was
derived for a nearly spherical potential, the loop orbits in triaxial
potentials also result in Mathieu-like equations for the instability
of the perpendicular coordinate (although the fraction of loop orbits
tends to decrease with increasing levels of triaxiality).  Note that
Mathieu's equation can also be transformed into a discrete map using
the same procedure as described in the previous subsections.

\section{CONCLUSION} 

\subsection{Summary of Dynamical Results} 

{\sl Analytic results:} One set of results from this paper is the
development of a collection of analytic expressions that describe
potentials, forces, and orbits for the triaxial generalization of the
Hernquist/NFW profile in the inner limit. In particular, we have found
a completely analytic form for the potential (eqs. [\ref{eq:phistart}
-- \ref{eq:finaltwo}]), the force terms (eqs. [\ref{eq:xforce} --
\ref{eq:zforce}]; see also PM01), the axis ratios of the potential
(eqs.  [\ref{eq:phiaxis} -- \ref{eq:jzint}]), and the potential in the
limit of nearly spherical density profiles (eqs. [\ref{eq:phidelt} --
\ref{eq:rhodelt}]). We have also found the radial solutions in the
spherical limit for the full Hernquist potential (\S 3.3), i.e., valid
over all radii $0 < \xi < \infty$, not just the inner limit $\xi \ll
1$. Finally, for the axisymmetric version of the NFW profile, we have
found analytic expressions for the potential and the force terms that
are valid over the entire radial range of the halo (see Appendix A).

Another contribution of this paper is an analytic formulation of the
instability problem. For orbits initially confined to one of the
principal planes, the equation of motion for the perpendicular
component takes a relatively simple form (eqs. [\ref{eq:omegax} --
\ref{eq:omegaz}]). Further, the instability can be described in terms
of simple model equations (\S 5, Appendices B -- D) that allow for an
analytic description of the instability. Since these equations of
motion have the form of Hill's equation, we can utilize a large
collection of known results (AS70, AW05, MW66). In this context,
however, the relevant Hill's equations have random forcing terms
(forcing strengths that vary from cycle to cycle), which requires the
development of new mathematical results. We have addressed this
complication in a separate companion paper (AB07).

In addition to their usefulness in providing physical understanding of
orbits and their instabilities, these analytic results are also useful
for computational purposes. Although all of these results can also be
determined numerically, the regime of parameter space is enormous, so
that analytic results provide a substantial savings in the required
computational effort. Analytic results also aid in our understanding
of the underlying mechanism for the instability, as outlined below.

{\sl Overview of the instability:} The main focus of this paper is to
demonstrate that for orbits initially confined to any of the principal
planes, the motion in the perpendicular direction is unstable to
growth; our related objective is to understand the underlying
mechanism for the instability. As discussed in \S 4, the instability
displays four key features: quasi-periodic oscillatory behavior,
exponential growth, superimposed chaotic variations, and eventual
saturation.  These four features can be understood as follows:

For a given energy, an orbit has a well-defined crossing time, which
naturally gives rise to quasi-periodic behavior. Furthermore, this
property provides an effective periodic forcing potential for the
motion in the direction perpendicular to the original orbital
plane. The resulting dynamics can thus be described with a
Mathieu-Hill-like equation, which is well-known to have unstable
(growing) solutions (MW66, AS70, AW05).  As a result, for a wide range
of parameters, the motion in the perpendicular direction will be
unstable. In the case of a ``pure'' Hill's equation, with precisely
periodic forcing, the growing solutions have exponential growth
superimposed on periodic oscillations, as seen in the instability
under study. The period of the growing solutions is given by the
period of the forcing function, which is determined here by the
crossing time of the orbit in the original plane.  Thus, the first two
properties of the instability (periodic behavior and exponential
growth) follow directly from the form of the equation of motion for
the perpendicular coordinate (eqs.  [\ref{eq:omegax} --
\ref{eq:omegaz}]).

Chaotic variations represent the next property of the instability and
they arise due to the triaxial nature of the potential (which arises
from the triaxial density profile).  As is well known (BT87), triaxial
potentials allow for box orbits, and for chaotic behavior, and both of
these effects are increasingly common as the axis ratios of the
triaxial potential become more extreme. For sufficiently triaxial
potentials, the crossing time varies from cycle to cycle and hence the
frequency of the forcing of the instability varies somewhat. However,
a much more significant effect is that the strength of the forcing
varies from cycle to cycle. In this case, the forcing strength is
determined by the distance of closest approach of the orbit to the
origin in the original plane, where this distance is weighted by the
axis weights $(a,b,c)$ of the triaxial potential.  For example, box
orbits are quasi-periodic and wander arbitrarily close to the origin.
Thus, as the orbit varies chaotically in its original plane, the
forcing terms for the instability vary as well, and these chaotic
variations appear in the growing (unstable) solution for the
perpendicular coordinate.

The final property of the unstable solutions is their eventual
saturation.  This behavior is expected for two reasons: In physical
terms, the full three dimensional orbit must conserve energy, so there
is a limit to the amplitude of oscillations in the perpendicular
(unstable) direction, i.e., saturation is inevitable. In mathematical
terms, the forcing strength depends (inversely) on the distance of
closet approach to the origin; for orbits with sufficiently large
values of the perpendicular coordinate, the forcing strength becomes
small and the instability stalls.

{\sl Model equations:} We have analyzed the instability of Hill's
equation in the delta function limit, where the forcing terms are
localized in time. In this limit, the equation of motion can be solved
analytically for a given cycle of the motion in the original plane.
As a result, the growth rates for a pure Hill's equation can also be
found analytically (\S 5 and Appendix B).  For the case of forcing
strengths that vary from cycle to cycle, we have found estimates for
the growth rates in terms of the expectation value of the growth rates
for individual cycles (Appendix C).  In the limit of highly unstable
systems, this expectation value -- the asymptotic growth rate --
provides a good working estimate. However, the true growth rate
contains an additional component (the anomalous growth rate) that
arises from the matching conditions at the cycle boundaries. In a
separate paper we show that this correction term can be found
analytically in terms of an expectation value (see AB07), and we
present general theorems regarding the instability and growth rates
for Hill's equation with random forcing terms.

{\sl The $\hat y$ direction is most unstable:} Although orbits around
all three axes are unstable, the orbits around the intermediate ($\hat
y$) axis are more likely to be unstable: The instability we are
exploring here is often strongest when the motion in the original
plane arises from box orbits, which are more likely to take place when
the axis ratios of the two axes in the original orbit plane are more
extreme (BT87). The ``$\hat y$ instability'' takes place when the
original orbit lies in the $x-z$ plane, and the axis ratio $a/c$ is
the largest by definition.  Therefore, the $\hat y$ instability is the
most likely to occur; alternatively, the growth rate for the $\hat y$
instability is likely to be larger than that for the other two cases.
One should compare with situation with the classic result from
rigid-body dynamics: As is well-known,\footnote{This calculation is
generally credited to Euler (1749) in classical mechanics texts (e.g.,
Marion 1970).}  rotation about the shortest or longest axis of an
asymmetric rigid body produces stable motion, whereas rotation about
the intermediate axis is unstable.

\subsection{Astrophysical Implications} 

Although this paper focuses on the description of the triaxial
potential, its orbits, and the corresponding orbital instabilities,
these results can be applied to a number of astrophysical systems on a
variety of size scales.  On the largest scales, this work applies to
the inner regions of dark matter halos -- both for galaxies and galaxy
clusters.  On smaller scales, these results are applicable to the
dynamics of tidal streams, galactic bulges and warps, and young
embedded star clusters.  All of these systems (or the inner portions
of them) are described by the triaxial potential developed herein, and
can be potentially affected by this orbit instability. These
applications are briefly outlined below, but we emphasize that they
should be explored in greater depth.

For completeness, we note that in real astrophysical systems, the
gravitational potential will never be completely smooth as assumed
here. Instead, stars and other inhomogeneities provide a nonzero level
of graininess to the potential, and will thereby contribute to the
degree of chaos in the constituent orbits. In terms of the instability
considered here, these irregularities guarantee that orbits cannot be
exactly confined to any given plane; instead, these effects act to set
the smallest possible amplitude for perturbations in the direction
perpendicular to the starting plane. We also note that the instability
considered here strictly applies only to orbits that start in one of
the principal planes, and such orbits represent only a small portion
of phase space.

{\sl Dark Matter Halos:} In dark matter halos, both the orbits of dark
matter particles themselves and the orbits of sub-halos benefit from
an analytic description, which can aid in understanding how dark
matter halos achieve the (nearly) universal form observed in numerical
simulations (starting with NFW). Another finding from numerical
simulations is that the velocity distributions are generally radial in
the outer parts of the halo and more isotropic in the inner
regions. This isotropization cannot take place through two-body
relaxation processes, as the time scale is much too long. The orbit
instability considered in this paper can provide at least a partial
explanation: The inner regions tend to have triaxial forms with
density profiles $\rho \sim 1/m$ in the inner limit, i.e., the form of
the density and potential considered in this paper. Since nearly
radial orbits are subject to the orbit instability considered herein,
the inner halo cannot maintain purely radial velocity
distributions. Further, both the time scale (several crossing times)
and the saturation level (the amplitudes of the perpendicular
coordinate oscillations become roughly comparable to those in the
original orbit plane) suggest that the velocity distribution can
become sufficiently isotropic to explain the results of numerical
simulations. The velocity distributions of dark matter particles, as
well as their enhancements in tidal streams (see below), can affect
direct detection strategies (e.g., Evans et al. 2000).

{\sl Tidal Streams:} Since tidal streams must orbit in galactic
potentials with some triaxial aspect, the results of this paper also
inform their dynamical properties. The coherence of tidal streams
relies on the orbits of the satellite galaxies and the stars that are
stripped away from them to remain coherent for a number of orbits. In
contrast, the instability studied in this paper can act to disrupt
tidal streams. This disruption could have important implications for
direct searches for dark matter, through particle detection
experiments, where enhancement due to tidal streams has been invoked
as a way to help discriminate the signal from the background (Evans et
al. 2000).  Turning the problem around, one can also use the observed
properties of tidal streams as a means to constrain the triaxiality of
their galactic potentials (Ibata et al. 2001). This instability will
also play a role in the assimilation of merging substructures during
the galaxy formation process (e.g., Helmi et al. 1999). In all of
these applications, however, the number of orbits of the satellites is
relatively small (several), and it could be hard to distinguish
between precession of an ordinary orbit in a triaxial potential and
this instability. Further exploration of this issues is necessary.

{\sl Galactic Bulges and Disks:} The orbits and instabilities studied
in this paper can also act on the scale of galactic bulges, which can
be described by a triaxial generalization of the Hernquist potential.
In this context, an important issue is how often stars wander close to
the potential center, where (usually) a supermassive black hole
resides. The orbit instability studied herein acts to change nearly
radial orbits into orbits with more isotropic velocity ellipsoids, and
thereby changes the rate at which stars interact with the central
black hole. We note that near the galactic center, the gravitational
potential of the black hole itself must be taken into account.  Some
work along these lines has been done (Gerhard \& Binney 1985; PM01,
Poon \& Merritt 2002), which shows that triaxiality leads to more box
orbits and greater rates of stellar accretion by a central black hole;
on the other hand, the orbit instability studied herein acts to
increase the distance of closest approach and thereby suppresses
accretion of stars. In any case, this issue provides an interesting
problem for future study.

A related issue is that sufficiently distorted (non-axisymmetric) disk
galaxies can be subject to this type of orbit instability, which acts
to populate regions out of the original disk plane.  This issue was
explored earlier by Binney (1981) using the triaxial version of the
logarithmic potential; this work can be used to extend such previous
analyses by providing a different potential and considering forcing
frequencies $\omega^2 (t)$ that have stochastic amplitude variations
from cycle to cycle and are highly nonlinear (e.g., closer to the
delta function limit than the cosine forcing functions that arise for
nearly spherical potentials -- see \S 5.3). Both gas and stars can be
subject to this instability in galactic disks, leading to galactic
warps and other interesting structure (e.g., Sparke 1995, 2002).

{\sl Embedded Star Clusters:} The dynamics of young embedded star
clusters can also be described using the inner limit of the Hernquist
potential (Larson 1985, Jijina et al. 1999, Adams et al. 2006).  These
clusters are expected to have more extreme axis ratios than the larger
scale systems considered above. Observations of embedded clusters
(Lada \& Lada 2003 and references therein) show that the youngest
systems are highly irregular, displaying large departures from
spherical symmetry. Somewhat older systems (a few Myr) are
significantly more spherical, so that the process of isotropization
takes place rapidly. In these systems, orbits can be altered through
star-star scattering events (where the stars generally have
accompanying disks and/or planets) and through the instability studied
in this paper.  If these young clusters form out of initial gas
configurations that are highly flattened, this instability will act to
populate the regions perpendicular to the initial cloud plane. The
resulting stellar clusters can thereby become more nearly spherical,
even in the absence of stellar scattering interactions.  For example,
suppose that the initial cluster is flattened with a 10:1 aspect ratio
($c$ = 0.1). For the orbit instability of this paper to round out the
cluster, the growth rate $\gamma$ must be large enough to compete with
stellar scattering. The number $N_C$ of crossing times required for
the instability to amplify the motion in the perpendicular direction
is given by $N_C \approx (\ln 10)/ \gamma t_c$.  For a purely stellar
system, the ratio of the dynamical relaxation time to the crossing
time (BT87) is given by $N_C \approx \nstar / (10 \ln \nstar)$; for an
embedded cluster dominated by its gas content, however, the relaxation
time is longer so that $N_C \approx (\nstar/\epsilon^2)/ [10 \ln
(\nstar/\epsilon)]$, where $\epsilon$ is the star formation efficiency
within the cluster (Adams \& Myers 2001). These relations imply that
orbit instability competes with dynamical relaxation when $\gamma t_c
> 10 \epsilon^2 \ln 10 \ln (\nstar/\epsilon) / \nstar$.  For typical
values ($\nstar$ = 300 and $\epsilon$ = 1/3), this constraint becomes
$\gamma t_c > 0.058$, a condition is that is often met. As a result,
orbit instabilities can dominate stellar scattering as a means to
isotropize young embedded star clusters.  In addition, the triaxial
nature of the potential in these clusters allows young star/disk
systems to execute box orbits, which bring the disks closer to the
massive stars at the cluster center. Triaxial potentials thus allow
for greater radiation exposure for star/disk systems, compared with
spherical clusters, or at least a wider distribution of radiative flux
experienced by individual solar systems.

\subsection{Discussion and Future Work} 

In addition to the applications outlined above, another issue is to
consider how this orbit instability affects the construction of
self-consistent galaxy models. In systems where the orbiting bodies
provide the source of the potential (e.g., stars in a galactic
setting) one must find self-consistent models where the orbiting stars
provide the gravity for the triaxial potential (Schwarzschild
1979). This issue is of particular importance for the intermediate
scale of galaxies and galactic bulges, and can be complicated by the
presence of supermassive black holes (e.g., Poon \& Merritt 2002). On
the larger scale of dark matter halos, the dark matter particles
provide the source (mass) for the gravitational potential. For these
systems, however, N-body simulations of large scale structure already
self-consistently take into account the orbits of the particles
(subject to finite numerical resolution) and are the motivation for
both the particular forms for the density profiles (NFW, Hernquist)
and their degree of triaxiality. Finally, on the smaller size scale of
embedded stellar clusters, the gravitational potential is dominated by
gas, which is subjected to additional forces (e.g., magnetic fields)
and does not execute the same orbits. As a result, in these systems
the background potential can be taken as fixed, at least for purposes
of studying this orbit instability.

The analysis of this paper is limited to the case of density profiles
that have the form $\rho \sim 1/r$ in the spherical limit, and this
form is applicable to only the inner regions of the extended mass
distributions in question. Future work should consider both the
generalization to other inner limiting forms (e.g., power-law density
profiles with other indices, $\rho \sim 1/m^\gamma$) and extension of
the potential out to all radii. Toward this latter goal, we present
the potential and force terms for the axisymmetric version of the NFW
profile in Appendix A. In addition to the study of other density
profiles, the available parameter space for the potentials considered
in this paper is enormous, and should be explored in greater depth.
For example, the case of resonant orbits, including their growth rates
for instability and saturation levels, would be particularly
interesting to consider (anonymous referee, private communication).

Finally, another direction for future work concerns the mathematics of
the instability mechanism. This paper presents a class of model
equations that describe the instability (\S 5), and our companion
paper proves basic Theorems regarding stochastic Hill's equations
(AB07), with a focus on the highly unstable limit. Future lines of
inquiry include finding analytic solutions for the case of forcing
terms with finite width, the relationship between lyapunov exponents
for chaotic orbits and the growth rates for unstable orbits, better
treatment of the saturation mechanism including predictions for the
saturation levels, and finding (and solving) analogous model equations
for generalized potentials. These mathematical issues, in conjunction
with the aforementioned astrophysical applications, provide a rich
collection of dynamical problems for further study.

\acknowledgements
We thank Charlie Doering, Gus Evrard, Lars Hernquist, Kelly
Holley-Bockelmann, and Tom Statler for useful discussions.  This work
was supported at the University of Michigan by the Michigan Center for
Theoretical Physics; by NASA through the Astrophysics Theory Program
(NNG04GK56G0) and the Spitzer Space Telescope Theoretical Research
Program (1290776); and by NSF through grants CMS-0408542 and
DMS-604307 (AMB).

\renewcommand{\theequation}{A\arabic{equation}}
\setcounter{equation}{0}  
\section*{Appendix A: Potential and Forces for Axisymmetric NFW Profile} 

In this Appendix, we present the potential and force terms for the
complete axisymmetric NFW profile.  The results are thus valid over
the entire range of radii, not just in the inner limit, but they are
limited to axial symmetry. 

Evaluating the potential integral for the oblate case, $c < 1$, 
yields the result, 
$$ 
\Phi (R,z) = {2 \over a - b} \Biggl\{ 
\sqrt{a - \gamma} \ln \Bigg| 
{ (\sqrt{a - \gamma} + c) (1 + \sqrt{1 - q/a}) ( \sqrt{1-q} - \sqrt{1 - q/a} ) \over 
(\sqrt{a - \gamma} - c) (1 - \sqrt{1 - q/a}) ( \sqrt{1-q} + \sqrt{1 - q/a} ) } \Bigg|  
$$
\be
+ 2 \sqrt{\gamma - b} \, \Bigl[ \tan^{-1} \bigl( 
{ \sqrt{\gamma - b} \over c } \bigr) \, - \, 
\tan^{-1} (q/b - 1)^{-1/2} + \tan^{-1} 
\bigl( { \sqrt{1-q} \over \sqrt{q/b - 1} } \bigr) 
\Bigr] \Biggr\} \, . 
\ee
Note that we have defined a number of new quantities, i.e., 
$$
\gamma \equiv 1 - c^2 \, \qquad \qquad \xi^2 \equiv R^2+z^2 \, 
\qquad \qquad q \equiv \gamma R^2 / \xi^2 \, , 
$$
\be
{\rm and} \qquad \qquad 
a,b = {1 \over 2} \left[ \gamma + \xi^2 \pm 
\sqrt{(\gamma + \xi^2)^2-4\gamma R^2} \right] \, . 
\ee
Keep in mind that $a$ and $b$ are the roots defined by equation (A2) and 
are not the axis weights as defined in the main text. 
For the corresponding prolate case, $c > 1$, the potential takes the form 
$$
\Phi (R,z) = {2 \over a - b} \Biggl\{ 
\sqrt{a - \gamma} 
\ln \Bigg|  
{(\sqrt{a - \gamma} + c) (1 + \sqrt{1 - q/a}) ( \sqrt{1-q} - \sqrt{1 - q/a} ) \over 
(\sqrt{a - \gamma} - c)(1 - \sqrt{1 - q/a}) ( \sqrt{1-q} + \sqrt{1 - q/a} ) } \Bigg|  
$$
\be 
- \sqrt{b-\gamma} \ln \Bigg| 
{(\sqrt{b - \gamma} + c) (1 + \sqrt{1 - q/b}) ( \sqrt{1-q} - \sqrt{1 - q/b} ) \over 
(\sqrt{b - \gamma} - c)(1 - \sqrt{1 - q/b}) ( \sqrt{1-q} + \sqrt{1 - q/b} ) } \Bigg|\Biggr\} \, . 
\ee
Note that the expressions for the prolate and oblate cases can be put
in similar form using the identity ${\rm tan}^{-1} x = (i/2) \ln \big|
(i+x)/(i-x) \big|$. Depending on the context, either form can be more 
computationally convenient. 

The forces are found by taking the negative gradient of the potential,
$F=-\nabla \Phi$.  Note that the potential is taken to be positive in
our convention, so that the positive gradient yields the forces. For
simplicity, we will present the result for the $\hat y$ component of
the force; the other forces can be found by simple substitution, as
outlined below.  The force for the oblate case is given by
$$
F_y = {{\partial a} \over {\partial y}} \Biggl\{ {2 \over (a-b)}
\Biggl[ {c \over (1-a)} + {\xi \over a} (1- {\sqrt{1-q} \over {1-a}}) \Biggr] +
{u(\gamma-\xi^2) \over \sqrt{a-\gamma} (a-b)^2 }  
- {4 \sqrt{\gamma-b} (\Theta-\delta+\eta) \over (a-b)^2} \Biggr\} +
$$
\be
{{\partial b}  \over {\partial y}} \Biggl\{ {2 \over (a-b)} 
\Biggl[ {c \over (b-1)} + {\xi \over b} \bigl( {\sqrt{1-q} \over {1-b}} - 1 \bigr) \Biggr]
+ {{2(\gamma-\xi^2)(\Theta-\delta+\eta)}\over {\sqrt{\gamma-b}(a-b)^2}} + 
{{2u \sqrt{a-\gamma}} \over (a-b)^2} \Biggr\} \, . 
\ee

\noindent 
The force for the prolate case is given by
$$
F_y={{\partial a}  \over {\partial y}}\Biggl\{{2 \over (a-b)}
\Biggl[ {c \over (1-a)}+{\xi \over a}(1-{\sqrt{1-q} \over {1-a}}) \Biggr] 
+ {u(\gamma-\xi^2) \over {\sqrt{a-\gamma} (a-b)^2}} 
+{2v\sqrt{b-\gamma}\over (a-b)^2}\Biggr\} +
$$
\be
{{\partial b}  \over {\partial y}} \Biggl\{{2 \over (a-b)} 
\Biggl[ {c \over (b-1)}+{\xi \over b}\bigl({\sqrt{1-q} \over {1-b}} -1\bigr) \Biggr]
+ {v\sqrt{\gamma-\xi^2}\over {\sqrt{b-\gamma}(a-b)^2}}+ 
{{2u \sqrt{a-\gamma}} \over (a-b)^2} \Biggr\} \, . 
\ee

\bigskip

The forces in the $\hat x$ and $\hat z$ directions are obtained by replacing the partial 
derivatives ${\partial a}  / {\partial y}$ and ${\partial b}  / {\partial y}$ with 
${\partial a}  / {\partial x}$ and ${\partial b}  / {\partial x}$ for the $\hat x$ 
direction, and ${\partial a}  / {\partial z}$ and ${\partial b} / {\partial z}$ 
for the $\hat z$ direction.

The above expressions have a number of new variables defined to make
them more manageable. The partial derivatives of the roots $a$ and $b$ 
are given by 
$$
{{\partial a}  \over {\partial x}} = x 
\Biggl\{ 1 + {{\xi^2-\gamma} \over \sqrt{(\gamma + \xi^2)^2-4\gamma R^2}} \Biggr\} 
= x \left[1 + {{\xi^2-\gamma} \over {(a-b)}} \right]\, , 
$$
\be
{{\partial b}  \over {\partial x}}=x 
\Biggl\{1 - {{\xi^2-\gamma} \over \sqrt{(\gamma + \xi^2)^2-4\gamma R^2}} \Biggr\} 
= x \left[1 - {{\xi^2-\gamma} \over {(a-b)}} \right]\, , 
\ee
\be
{{\partial a}  \over {\partial y}}=y \left[1 + {{\xi^2-\gamma} \over {(a-b)}} \right] 
\qquad \qquad {\rm and} \qquad \qquad 
{{\partial b}  \over {\partial y}}= y \left[1 - {{\xi^2-\gamma} \over {(a-b)}} \right]\, , 
\ee
\be
{{\partial a}  \over {\partial z}}= z \left[1 + {{\xi^2+\gamma} \over {(a-b)}} \right] \qquad
\qquad {\rm and} \qquad \qquad 
{{\partial b}  \over {\partial z}}=z \left[1 - {{\xi^2+\gamma} \over {(a-b)}} \right]\, .
\ee
We have also defined the quantities 
\be
\Theta \equiv \tan^{-1}\left({\sqrt{\gamma-b} \over c} \right) \, , \qquad 
\delta \equiv \tan^{-1}\left({1 \over \sqrt{q/b-1}}\right) \, , \qquad 
\eta \equiv \tan^{-1}\left( {\sqrt{1-q} \over \sqrt{q/b-1}} \right) \, , 
\ee
\be
u \equiv \ln \Bigg| { (\sqrt{a - \gamma} + c) (1 + \sqrt{1 - q/a}) 
( \sqrt{1-q} - \sqrt{1 - q/a} ) \over (\sqrt{a - \gamma} - c)
(1 - \sqrt{1 - q/a}) ( \sqrt{1-q} + \sqrt{1 - q/a} ) } \Bigg| \, , 
\ee
and finally 
\be
v \equiv \ln \Bigg| 
{(\sqrt{b - \gamma} + c) \over (\sqrt{b - \gamma} - c) } 
{ (1 + \sqrt{1 - q/b}) ( \sqrt{1-q} - \sqrt{1 - q/b} ) \over 
(1 - \sqrt{1 - q/b}) ( \sqrt{1-q} + \sqrt{1 - q/b} ) } \Bigg| \, . 
\ee

\renewcommand{\theequation}{B\arabic{equation}}
\setcounter{equation}{0}  
\section*{Appendix B: Solutions for the Delta Function Limit} 

This Appendix constructs the solutions for the instability equation in
the delta function limit (see eq. [\ref{eq:eqdelta}]) and finds the
growth rate as a function of the input parameters. In this limit, the
equation of motion has two linearly independent solutions $y_1(t)$ and
$y_2(t)$, which are defined through their initial conditions 
\be 
y_1 (0) = 1, \quad {dy_1 \over dt} (0)   = 0 , \qquad {\rm and} \qquad 
y_2 (0) = 0, \quad {dy_2 \over dt} (0) = 1 \, . 
\label{eq:bc} 
\ee 
The first solution $y_1$ has the generic form 
\be 
y_1 (t) = \cos \sqrt{\af} t 
\qquad {\rm for} \quad 0 \le t < \pi/ 2 \, , 
\ee
and 
\be
y_1 (t) = A \cos \sqrt{\af} t + B \sin \sqrt{\af} t 
\qquad {\rm for} \quad \pi/ 2 < t \le \pi \, , 
\ee
where $A$ and $B$ are constants that are determined by matching the
solutions across the delta function at $t = \pi/2$, including the
usual jump condition in the time derivative.  We define 
$\theta \equiv \sqrt{\af} \pi/2$ and find
\be 
A = 1 + (q/\sqrt{\af}) \sin\theta \cos\theta 
\qquad {\rm and} \qquad 
B = - (q/\sqrt{\af}) \cos^2 \theta \, . 
\ee
Similarly, the second solution $y_2$ has the form 
\be
y_2 (t) = \sin \sqrt{\af} t \qquad 
{\rm for} \quad 0 < t < \pi/2 \, , 
\ee
and 
\be 
y_2 (t) = C \cos \sqrt{\af} t + D \sin \sqrt{\af} t 
\qquad {\rm for} \quad \pi/ 2 < t \le \pi \, , 
\ee
where 
\be 
C = (q/\af) \sin^2 \theta \qquad {\rm and} \qquad 
D = {1 \over \sqrt{\af} } - (q/\af) \sin\theta \cos\theta 
\, . 
\ee 

Using the results derived thus far, we can find the criterion for
instability and the growth rate for unstable solutions.  The growth
factor $f_c$ per cycle is given by the solution to the characteristic
equation (see MW66) and can be written as
\be 
f_c = { \discrim + (\discrim^2 - 4)^{1/2} \over 2} \, , 
\ee
where the discriminant $\discrim = \discrim (\lambda, q)$ is defined by 
\be 
\discrim \equiv y_1 (\pi) + {dy_2 \over dt}(\pi) \, . 
\ee 
It then follows from Floquet's Theorem (see MW66, AS70) that
$|\discrim| > 2$ is a sufficient condition for instability. In
addition, periodic solutions exist when $|\discrim|$ = 2. At the
boundary between stability and instability, the parameter $\discrim$ 
= 2.  Solutions with $\discrim$ = 2 are usually unstable, but not 
always, so that further analysis is necessary for this case (MW66).  
Since the forcing potential is symmetric, we know that $y_1 (\pi) =
{\dot y}_2 (\pi)$, from Theorem 1.1 of MW66. If we define $H \equiv
|y_1 (\pi)| = | {\dot y}_2 (\pi)|$, the resulting criterion for
instability reduces to the form
\be 
H = \Bigg| {q \over 2 \sqrt{\af} } \sin (\sqrt{\af} \pi) - 
\cos (\sqrt{\af} \pi) \Bigg| > 1 \, , 
\label{eq:hdef} 
\ee 
and the corresponding growth rate $\gamma$ is given by 
\be 
\gamma = {1 \over \pi} \ln [ H + \sqrt{H^2 - 1} ] \, .
\ee

\renewcommand{\theequation}{C\arabic{equation}}
\setcounter{equation}{0}  
\section*{Appendix C: Random Variations in Forcing Strength} 

The analysis presented in \S 5 considers the forcing function $Q(t)$
to be perfectly periodic. However, the physical orbits found in a
triaxial potential often have a random element so that the amplitude
$q$ of the function $Q(t)$ varies from cycle to cycle. This Appendix
addresses this random variation by considering one cycle at a time.
Specifically, we consider each period from $t=0$ to $t=\pi$ as a
separate cycle, and then consider the effects of successive cycles
with varying values of forcing strength $q$.

During any given cycle (for any value of forcing strength $q_j$), the
solution can be written as a linear combination of $y_1$ and $y_2$.
Consider two successive cycles. The first cycle has forcing strength
$q_a$ and solution
\be 
f_a (t) = \alpha_a y_{1a} (t) + \beta_a y_{2a} (t) \, , 
\ee 
where the solutions $y_{1a}(t)$ and $y_{2a}(t)$ correspond to those 
found in the previous subsection when evaluated using the value $q_a$. 
Similarly, for the second cycle with $q$ = $q_b$, the solution has 
the form
\be 
f_b (t) = \alpha_b y_{1b} (t) + \beta_b y_{2b} (t) \, .  
\ee 
Next we note that the new coefficients $\alpha_b$ and $\beta_b$ are 
related to those of the previous cycle through the relations 
\be
\alpha_b = \alpha_a y_{1a} (\pi) + \beta_a y_{2a} (\pi) 
\qquad {\rm and} \qquad \beta_b = 
\alpha_a {d y_{1a} \over dt} (\pi) + 
\beta_a {d y_{2a} \over dt} (\pi) \, . 
\ee
The new coefficients can thus be considered as a two dimensional
vector, and the transformation between the coefficients in one cycle
and the next cycle is a $2 \times 2$ matrix. In addition, since we
have the analytic solution for any given cycle from the previous
subsection, the coefficients of this matrix are known. Finally, since
the equation is symmetric with respect to the midpoint $t = \pi/2$,
and since the Wronksian of the original differential equation 
(\ref{eq:modeleq}) is unity, the number of independent matrix 
coefficients is reduced from four to two. The transformation can 
thus be written in the form 
\be 
\left[ \matrix{ \alpha_b \cr \beta_b \cr } \right] = 
\left[ \matrix{ h & (h^2 -1)/g \cr g & h \cr } \right] 
\left[ \matrix{ \alpha_a \cr \beta_a \cr } \right] \equiv {\bf M} (q_a) 
\left[ \matrix{ \alpha_a \cr \beta_a \cr } \right] \, , 
\label{eq:matrixdef} 
\ee
where the matrix $\bf M$ (defined in the second equality) depends
on the value of $q_a$, and where the matrix coefficients are given by
\be
h = \cos (\sqrt{\lambda}\pi) - {q \over 2 \sqrt{\lambda} } 
\sin (\sqrt{\lambda}\pi) 
\qquad {\rm and} \qquad 
g = - \sqrt{\lambda} \sin (\sqrt{\lambda}\pi) - 
q \cos^2 (\sqrt{\lambda}\pi/2) \, . 
\label{eq:elements} 
\ee
After $N$ cycles with varying values of $q = q_k$ (where $k$ ranges 
from 1 to $N$), the solution retains the general form given above
where the coefficients are determined by the product of matrices 
according to
\be 
\left[ \matrix{ \alpha_N \cr \beta_N \cr } \right] = {\bf M}^{(N)} 
\left[ \matrix{ \alpha_0 \cr \beta_0 \cr } \right] 
\qquad {\rm where} \qquad 
{\bf M}^{(N)} \equiv \prod_{k=1}^N {\bf M}_k (q_k) \, . 
\label{eq:product} 
\ee
With this solution, we have transformed the original differential
equation (with a random element) into a discrete map. Further, the
properties of the product matrix ${\bf M}^{(N)}$ determine whether or
not the solution is unstable and the corresponding growth rate. 

The evaluation of the growth rate thus requires the repeated
multiplication of $2\times2$ matrices with randomly sampled $q_j$
values, where the matrix elements are defined through equations
(\ref{eq:matrixdef}) and (\ref{eq:elements}).  An estimate for the
growth rate can be found by considering each cycle to experience the
maximum possible growth, given the parameters of that cycle.  Under
this assumption, the growth rate -- denoted here as the asymptotic
growth rate $\gamma_\infty$ -- is given by
\be 
\gamma_\infty = {1 \over \pi N} \ln \left[ 
\prod_{k=1}^N \bigl\{ H_k + \sqrt{H_k^2 - 1} \bigr\} \right]  \, , 
\label{eq:gmax} 
\ee
where $H_k = H(q_k)$ is given by equation (\ref{eq:hdef}) evaluated at
$q=q_k$. It is understood that if $|H_k| < 1$ for a particular cycle,
then the growth factor is unity for that cycle.  Notice that 
this expression reduces to the form 
\be 
\gamma_\infty = {1 \over \pi N} 
\sum_{k=1}^N \ln \bigl\{ H_k + \sqrt{H_k^2 - 1} \bigr\} \, = 
{1 \over N} \sum_{k=1}^N \gamma_k = \langle \gamma \rangle \, .  
\label{eq:gmaxtwo} 
\ee
In other words, the asymptotic growth rate is the expectation value of
the particular growth rates for the individual cycles.

In practice, however, the growth rates of the solutions will not be
exactly equal to $\gamma_\infty$.  During each cycle, the solution
must contain an admixture of both the growing solution and the
decaying solution over that cycle.  Further, the solutions must match
up at the cycle boundaries. Counterintuitively, this effect tends to
make the true growth rates somewhat larger than the asymptotic growth
rate defined here (see AB07 for greater detail).  Numerical
experimentation indicates that the asymptotic growth rate (as defined
in eq.  [\ref{eq:gmax}]) is generally within 20\% of the true value,
although this result varies with the part of parameter space under
consideration. The analysis of AB07 shows that both the variations in
$\lambda$ and the matching conditions across cycle boundaries can
produce significant contributions to the growth rate. This correction
is important for cases in which the variance $\sigma^2$ of the ratios
of the matrix elements becomes extremely large, a limit that is not
expected to apply to astrophysical orbit problems such as those
considered here. Specifically, this correction to the growth rate has
the form $\Delta \gamma = \sigma^2 /(8 \pi)$ in the limit $\sigma \ll
1$ and the form $\Delta \gamma = C_\infty \sigma$ in the limit $\sigma
\gg 1$, where the constant $C_\infty$ depends on the shape of the
distribution (AB07).

\renewcommand{\theequation}{D\arabic{equation}}
\setcounter{equation}{0}  
\section*{Appendix D: Generalization to Aperiodic Variations} 

Most of the discussion in the main text (\S 5) considers the forcing
function $Q(t)$ to have zero width (the delta function limit) and to
be exactly $\pi$-periodic. Using the development of a discrete map
found in Appendix C, we can relax the second assumption (the first
assumption is generalized in \S 5.2).  As before, we consider one
cycle at a time, where each cycle has an amplitude $q_j$, but the
period is now given by $\mu_j \pi$.  In the development thus far, we
have defined the time variable to make the period equal to $\pi$
(which can always be done as long as the period is constant). In this
generalized treatment, we define the time variable so that the mean
period is $\pi$. As a result, the average of $1/\mu_j$, which we
denote with angular brackets, is taken to be unity so that $\langle
1/\mu_j \rangle$ = 1.

The equation of motion now takes the form 
\be
{d^2 y \over dt^2} + 
\left[ \lambda_j + q_j {\hat Q} (\mu_j t) \right] y = 0 \, , 
\ee 
where we have normalized the forcing frequency to have unit amplitude
($\hat Q$ = $Q/q_j$). Since $\hat Q$ (and $Q$) are $\pi$-periodic, the
$jth$ cycle is defined over the time interval $0 \le \mu_j t \le \pi$,
or $0 \le t \le \pi/\mu_j$. For each cycle, we can thus re-scale both 
the time variable and the ``constants'' according to 
\be 
t \to \mu_j t \, , \qquad 
\lambda_j \to \lambda_j / \mu_j^2 = \lamwig \, , \qquad 
{\rm and} \qquad q_j \to q_j/\mu_j^2 = \qwig \, ,  
\label{eq:rescale} 
\ee 
so the equation of motion reduces to the familiar form 
(see Theorem 1 of AB07)
\be 
{d^2 y \over dt^2} + 
\left[ \lamwig + \qwig {\hat Q} (t) \right] y = 0 \, . 
\ee 
For example, if we consider the delta function limit, then ${\hat Q}
(t) = \delta (\tmod - \pi/2)$, and the solution for a given cycle is
determined by the matrix ${\bf M} (q_j)$, which is defined by equation
(\ref{eq:matrixdef}) evaluated using $\lamwig$ and $\qwig$ to specify
the matrix elements. Notice also that we have allowed the parameter
$\lambda_j$ to also vary from cycle to cycle within this formalism.
Since we need to re-scale the value of $\lambda_j$ every cycle to
account for the period variation (specified by $\mu_j$), the net cost
of evaluation is the same. The physical implication of this result is
that departures from exact periodicity (in the time interval, not the
amplitude) lead to an additional effective variation in the forcing
strength $q_j$ and the base frequency (determined by $\lambda_j$), but
do not otherwise affect the instability. In general, an increased
variance in the distribution of the parameters $(\lambda_j,q_j)$ leads
to larger growth rates (AB07).

For the orbits considered in this paper, the period does not vary from
cycle to cycle as much as the forcing strength (see Fig. \ref{fig:qphist}). 
For example, the standard deviation $\sigma_\mu$ of the relative
(scaled) period is generally small compared to unity, so that the
parameter $\mu_j$ typically lies in the range given by $\mu_j = 1 \pm
\sigma_\mu$, where $\sigma_\mu \ll 1$.  As a result, the corrections
to the frequency parameter and forcing strength parameter (from
eq. [\ref{eq:rescale}]) have a typical relative size of only $2
\sigma_\mu \ll 1 $. The variations in $q_j$ from cycle to cycle are
much larger and dominate the random component of the equation of
motion for the instability.

\renewcommand{\theequation}{D\arabic{equation}}
\setcounter{equation}{0}  
\section*{Appendix E: Heuristic Discussion of Hill's Equation} 

In this Appendix we outline the essential features of Hill's equation,
the reasons for instability, and the relation of these issues to the
solutions found in this paper.  In rough terms, the form of Hill's
equation (or, equivalently, Mathieu's equation) allows for unstable
solutions because it acts like a driven oscillator. We first note that
if the forcing term were absent ($Q(t)$ = 0), then the equation would
describe simple harmonic motion with a frequency given by
$\sqrt{\lambda}$. For nonzero forcing functions, one can think of the
forcing term as being on the right hand side of the equation, so that
one obtains a driven oscillator, which can have growing (unstable)
behavior. In this case, the driving term is complicated by the fact
that it depends on the solution, $y(t)$, which in turn depends on both
the driving term and the natural oscillation frequency
$\sqrt{\lambda}$.  However, this complication does not prevent
unstable growth.

Next we note that since there are two frequencies involved, the
periodic driving frequency and the natural oscillation frequency, the
two effects can interfere either constructively or destructively. As a
rough illustration of this behavior, one can imagine pushing a child
on a swing (e.g., Arnold 1978).  If the pushes are in phase with the
oscillatory motion, the swing gains amplitude; if the pushes are out
of phase, then the swing loses amplitude and the oscillations are
damped. Because of this tuning, or phase requirement, Hill's equation
has regions of both stability and instability, depending on the
relative phases of the two frequencies involved. As result, the plane
of parameters shows alternating stripes of stability and instability,
as shown by Figures \ref{fig:qlplane} and \ref{fig:qlplaneb}.  In the
delta function limit, we can find analytic solutions for a given set
of parameters (Appendix B), or for a given cycle in the case of random
variations (Appendix C).  The corresponding solutions, or the matrix
elements, can be written in terms of sines and cosines of the angle
$\theta \equiv \sqrt{\lambda} \pi$. Here, $\sqrt\lambda$ is the
natural oscillation frequency and $\pi$ is the period (by
construction) of the forcing term. The combination of these two
quantities thus determines the solutions and the growth rates. As the
frequency $\sqrt{\lambda}$ varies (for fixed period $\pi$ of the
driving term), the system goes in and out phase, resulting in the
alternating bands of stability and instability shown in the figures.

The tuning described above depends on the amplitude $q$ of the driving
term. Much like for the case of pushing a swing, a stronger driving
term allows for greater error (leeway) in the phase matching necessary
to maintain unstable growth. Since the required tuning to obtain
instability becomes less delicate, more of the available parameter
space allows for instability, and the regions of stability get smaller
with increasing forcing amplitude $q$ (Figs. \ref{fig:qlplane} and
\ref{fig:qlplaneb}).  Nonetheless, the regions of stability never
vanish completely. The analysis of Appendices B and C illustrates this
behavior quantitatively for the simplest case of Hill's equation in
the delta function limit.

\end{document}